\documentclass[aps,prb,groupedaddress,showpacs,twocolumn,floatfix]{revtex4}
\usepackage{graphicx}
\usepackage{amsmath}
\usepackage{bbm}
\usepackage{xspace}
\usepackage{color}

\newcommand{\se}{Sec.\@\xspace}

\newcommand{\app}{App.\@\xspace}
\newcommand{\ie}{i.\thinspace{}e.\@\xspace}

\newcommand{\ptl}{\partial}

\newcommand{\PDF}[2]{\frac{\ptl\, #1}{\ptl\, #2}}

\newcommand{\ve}[1]{{\bf #1}}
\newcommand{\mat}[1]{\mathsf{#1}}

\newcommand{\eq}[1]{Eq.\thinspace{}(\ref{#1})}
\newcommand{\eqq}[2]{Eqs.\thinspace{}(\ref{#1}) and (\ref{#2})}
\newcommand{\Eq}[1]{Equation (\ref{#1})}

\newcommand{\tab}[1]{Tab.\thinspace{}\ref{#1}}

\newcommand{\fig}[1]{Fig.\thinspace{}\ref{#1}}
\newcommand{\figg}[2]{Figs.\thinspace{}\ref{#1} and \ref{#2}}
\newcommand{\fc}[1]{({#1})}
\newcommand{\figc}[2]{Fig.\thinspace{}\ref{#1}\thinspace{}\fc{#2}}

\newcommand{\Fig}[1]{Figure \ref{#1}}

\newcommand{\Figc}[2]{Figure \ref{#1}\thinspace{}\fc{#2}}

\newcommand{\Tr}{\mbox{Tr}}
\newcommand{\tr}{\mbox{tr}}

\def\bra#1{\mathinner{\langle{#1}|}}
\def\ket#1{\mathinner{|{#1}\rangle}}
\def\braket#1{\mathinner{\langle{#1}\rangle}}

\DeclareMathOperator{\Qmat}{Q}
\newcommand{\Qp}[1]{{\sideset{^p}{_{#1}}\Qmat}}
\newcommand{\Qh}[1]{{\sideset{^h}{_{#1}}\Qmat}}
\newcommand{\Qpdag}[1]{{\sideset{^p}{^\dagger_{#1}}\Qmat}}
\newcommand{\Qhdag}[1]{{\sideset{^h}{^\dagger_{#1}}\Qmat}}

\DeclareMathOperator{\diag}{diag}

\begin{document}

% Use the \preprint command to place your local institutional report
% number in the upper righthand corner of the title page in preprint mode.
% Multiple \preprint commands are allowed.
% Use the 'preprintnumbers' class option to override journal defaults
% to display numbers if necessary
%\preprint{}

%Title of paper
\title{Spectral properties of coupled cavity arrays in one dimension}

% repeat the \author .. \affiliation  etc. as needed
% \email, \thanks, \homepage, \altaffiliation all apply to the current
% author. Explanatory text should go in the []'s, actual e-mail
% address or url should go in the {}'s for \email and \homepage.
% Please use the appropriate macro foreach each type of information

% \affiliation command applies to all authors since the last
% \affiliation command. The \affiliation command should follow the
% other information
% \affiliation can be followed by \email, \homepage, \thanks as well.
\author{Michael Knap}
\email[]{michael.knap@tugraz.at}
\affiliation{Institute of Theoretical and Computational Physics, Graz University of Technology, 8010 Graz, Austria}
\author{Enrico Arrigoni}
\affiliation{Institute of Theoretical and Computational Physics, Graz University of Technology, 8010 Graz, Austria}
\author{Wolfgang von der Linden}
\affiliation{Institute of Theoretical and Computational Physics, Graz University of Technology, 8010 Graz, Austria}
%\homepage[]{Your web page}
%\thanks{}
%\altaffiliation{}

%Collaboration name if desired (requires use of superscriptaddress
%option in \documentclass). \noaffiliation is required (may also be
%used with the \author command).
%\collaboration can be followed by \email, \homepage, \thanks as well.
%\collaboration{}
%\noaffiliation

\date{\today}

\begin{abstract}
Spectral properties of coupled cavity arrays in one dimension are investigated by means of the variational cluster approach. Coupled cavity arrays consist of two distinct ``particles,'' namely, photons and atomiclike excitations. Spectral functions are evaluated and discussed for both particle types. In addition, densities of states, momentum distributions and spatial correlation functions are presented. Based on this information, polariton ``quasiparticles'' are introduced as appropriate, wave vector and filling dependent linear combinations of  photon and atomiclike particles. Spectral functions and densities of states are evaluated for the polariton quasiparticles, and the weights of their components are analyzed.
\end{abstract}

% insert suggested PACS numbers in braces on next line
%64.70.Tg: Quantum phase transitions (for quantum Hall effects aspects, see 73.43.Nq in electronic structure of surfaces, interfaces, thin films, and low dimensional structures)
%67.85.De: Dynamic properties of condensates; excitations, and superfluid flow
%03.75.Kk: Dynamic properties of condensates; collective and hydrodynamic excitations, superfluid flow
%71.36.+c: Polaritons (including photon-phonon and photon-magnon interactions)
%73.43.Nq       Quantum phase transitions (see also 64.70.Tg Quantum phase transitions in equations of state, phase equilibria and phase transitions)
%64.70.-p       Specific phase transitions
%42.50.Ct       Quantum description of interaction of light and matter; related experiments
% insert suggested keywords - APS authors don't need to do this
%\keywords{}

%\maketitle must follow title, authors, abstract, \pacs, and \keywords
\maketitle

% body of paper here - Use proper section commands
% References should be done using the \cite, \ref, and \label commands
\section{\label{sec:introduction}Introduction}
\begin{figure}
        \centering
        \includegraphics[width=0.48\textwidth]{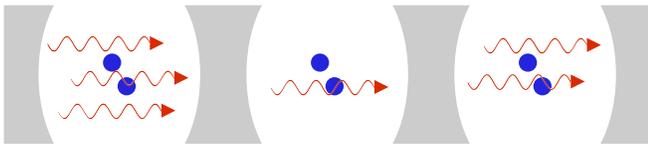}
        \caption{Cavities forming a one-dimensional chain lattice. The blue dots represent atomic systems, whereas the red wavy arrows indicate photons. }
        \label{fig:IntCav}
\end{figure}

The experimental progress in controlling quantum optical and atomic
systems, which has been achieved over the last few years, 
prompted
ideas for new realizations of strongly correlated many body systems,
such as ultracold gases of atoms trapped in optical lattices
\cite{jaksch_cold_1998, greiner_quantum_2002, bloch_many-body_2008} or
light-matter systems.\cite{greentree_quantum_2006,
  hartmann_strongly_2006, hartmann_quantum_2008} The latter consist of
photons, which 
interact
with atoms or atomiclike
structures. Normally, the interaction between photons and atoms is very
weak, since the interaction time is small. However, a strong
interaction can be achieved when photons are confined within optical
cavities.
In this case, the
coupling between photons and atoms leads to 
an effective 
repulsion between
photons, which means that it costs energy to add additional photons to
the cavity. The arrangement of such cavities on a lattice, see
\fig{fig:IntCav}, allows the photons to ``hop'' between neighboring
sites, provided the cavities are coupled.
Quantum mechanically the coupling of adjacent cavities means that
their photonic wave functions overlap. Due to the strong interaction
between photons and atoms, and the introduction of a lattice of
coupled cavities, a strongly correlated phase emerges where photons
are present. The light-matter models share some basic properties with
the Bose-Hubbard (BH) model, \cite{fisher_boson_1989} such as the
quantum phase transition from a Mott phase, where particles are
localized on the lattice sites, to a superfluid phase, where particles
are delocalized on the whole lattice. \cite{greentree_quantum_2006}
Yet the physics of the light-matter models is far richer because two
distinct particles, namely, photons and atomiclike excitations, are present.

A major advantage of these man-made realizations of
  strongly correlated many-body systems is that they 
can be tailored to correspond to
a many-body model, whose
parameters can be directly controlled in the experiment. Furthermore local quantities, such as the particle density at a specific lattice site, can be addressed individually due to the mesoscopic scale of the cavities and both lattice size and geometry can be controlled in the fabrication process.
An experimental realization of these light-matter systems is still
missing but there are several promising approaches, such as photonic
crystal cavities or toroidal and disk-shaped
cavities. \cite{hartmann_quantum_2008} If light-matter systems can be
realized, they will undoubtedly provide fascinating insight in the
physics of strongly correlated many-body systems. The realizations
might be used as quantum simulators for other quantum mechanical
problems or even more intriguing for quantum information processing
applications.\cite{illuminati_quantum_2006}

Recently, there has been a lot of research activity in the field of
light-matter systems. Most of the work has been devoted to investigate
the quantum phase transition from the Mott to the superfluid phase. Some
basic characteristics of the quantum phase transition have been
evaluated from small systems of a few cavities by means of exact
diagonalization.\cite{hartmann_strongly_2006, hartmann_strong_2007,
  angelakis_photon-blockade-induced_2007, makin_quantum_2008,
  irish_polaritonic_2008} Results are available at mean-field level
\cite{greentree_quantum_2006, koch_superfluid--mott-insulator_2009,
  lei_quantum_2008, na_strongly_2008} as well or more accurately from
analytical strong coupling perturbation theory
calculations,\cite{schmidt_strong_2009} and from simulations based on
the density matrix renormalization
group\cite{rossini_mott-insulating_2007, rossini_photon_2008} (DMRG),
the variational cluster approach\cite{aichhorn_quantum_2008} and
Quantum Monte Carlo.\cite{zhao_insulator_2008} Spectral properties of
light-matter systems have been investigated in
Refs. \onlinecite{schmidt_strong_2009},
\onlinecite{aichhorn_quantum_2008}, and
\onlinecite{pippan_excitation_2009}.

In the present paper, we study in detail the spectral properties of a
one-dimensional light-matter system.
In particular,
we evaluate both photonic as well as atomic-excitation spectral
functions. 
The investigation of both spectral functions allows us to
characterize the 
polariton 
excitations in light-matter models. In
addition to the spectral functions, we present densities of states,
momentum distributions and spatial correlation functions. For
completeness we also show the first two lobes delimiting  the Mott
transition.

This paper is organized as follows: in \se~\ref{sec:model}, we
introduce the light-matter model. Section~\ref{sec:method} contains both the
description of the numerical method as well as the exploration of the polaritonic properties. Section~\ref{sec:results}
is devoted to spectral properties of the light-matter system. Here, we
present our results for spectral functions, densities of states, momentum
distributions and spatial correlation functions. Finally, we summarize
and conclude our findings in \se~\ref{sec:conclusion}.

\section{\label{sec:model}Model}
From the great variety of possible theoretical descriptions of
light-matter systems \cite{greentree_quantum_2006,
  hartmann_strongly_2006, hartmann_quantum_2008,
  rossini_mott-insulating_2007, brandao_light-shift-induced_2008,
  ji_quantum_2007} we concentrate on the simplest one, which consists
of an array of cavities each of which contains a two-level
system.\cite{greentree_quantum_2006} The physics of the $i$-th cavity
can be described by the Jaynes-Cummings (JC) Hamiltonian,
\cite{jaynes_comparison_1963} which for $\hbar=1$ is given by
\begin{equation}
 \hat{H}^{JC}_i = \omega_c \, a_i^\dagger \, a_i + \epsilon \, \sigma_i^+ \, \sigma_i^-+g\left( a_i\,\sigma_i^+ + a_i^{\dagger}\,\sigma_i^- \right) \;\mbox{,}
 \label{eq:jc}
\end{equation}
where $\omega_c$ is the resonance frequency of the cavity, \ie, the
frequency of the confined photons, $\epsilon$ is the energy spacing of
the two-level system, and $g$ is the 
atom-field 
coupling constant. The
operator $a_i^\dagger$ creates a photon with frequency 
$\omega_c$,
whereas $a_i$ annihilates one. The two-level system can be
mathematically described by
Pauli spin algebra. Thus, we identify the
ground state of the two-level system with $\ket{\downarrow_i}$ and the
excited state with $\ket{\uparrow_i}$. With that the atomic raising
operator is defined as $\sigma_i^+ \equiv
\ket{\uparrow_i}\bra{\downarrow_i}$ and the atomic lowering operator
as $\sigma_i^- \equiv \ket{\downarrow_i}\bra{\uparrow_i}$,
respectively. In order to obtain the JC Hamiltonian the rotating wave
approximation, which is justified for $|\omega_c-\epsilon| \ll
\omega_c,\,\epsilon $,\cite{haroche_exploringquantum:_2006} has been
assumed. The deviation between the resonance frequency and the energy
spacing of the 
two-level system, $\Delta \equiv \omega_c - \epsilon$,
is termed detuning.
For the JC Hamiltonian the particle number
$\hat{n}_{i} = a_i^\dagger \, a_i + \sigma_i^+ \, \sigma_i^-$ is a
conserved quantity, as $[\hat{H}_i^{JC}, \, \hat{n}_{i}] = 0$. This is
a consequence of the rotating wave
approximation.\cite{haroche_exploringquantum:_2006}

The full model consists of an array of $N$ cavities, which form a
lattice and hence we refer to this model as the Jaynes-Cummings
lattice (JCL) model. Due to the coupling of the cavities, photons are
allowed to hop between neighboring lattice sites. This leads to the
JCL Hamiltonian
\begin{equation}
 \hat{H}^{JCL} = -t \sum_{\left\langle i,\,j \right\rangle} a_i^\dagger \, a_j + \sum_i \hat{H}_i^{JC} - \mu\,\hat{N}_p \;\mbox{,}
 \label{eq:jcl}
\end{equation}
where $t$ is the hopping strength and $\mu$ the chemical potential,
which controls the total particle number $\hat{N}_p$ of the
system. The first sum with the angle brackets around the summation indices
is restricted to nearest-neighbor sites. 
In the case of the JCL model, the particle number
of a specific cavity $\hat{n}_{i}$ is not conserved anymore. However,
the total particle number $\hat{N}_p=\sum_i \hat{n}_{i}$ is a conserved
quantity.
In summary, the 
JCL Hamiltonian can be rewritten as
\begin{align}
 \hat{H}^{JCL} = &-t \sum_{\left\langle i,\,j \right\rangle} a_i^\dagger \, a_j -\Delta \sum_i \sigma_i^+ \, \sigma_i^-  \nonumber \\
&+g\sum_i  ( a_i\,\sigma_i^+  + a_i^\dagger\,\sigma_i^- ) - (\mu-\omega_c)\,\hat{N}_p\;\mbox{.}
 \label{eq:spe:jclrewritten}
\end{align}
From \eq{eq:spe:jclrewritten} and from the fact that we consider the
coupling strength $g$ as unit of energy, it follows that
the physics only depends on three independent parameters,
namely the
hopping strength $t$, the detuning $\Delta$ and the modified chemical
potential $\mu-\omega_c$.
In order to fulfill the condition for the rotating wave approximation
the resonance frequency $\omega_c$ has to be large in comparison with
the detuning $\Delta$, which can be always satisfied theoretically as
solely the difference between the chemical potential and the resonance
frequency appears in the 
grand-canonical
Hamiltonian $\hat{H}^{JCL}$.

\section{\label{sec:method}Method}
In order to investigate the properties of the JCL model, we use the variational cluster
approach \cite{potthoff_variational_2003} (VCA), which has been
formulated for bosonic systems in
Ref.~\onlinecite{koller_variational_2006}.
Previous work on the JCL model within VCA was carried out in
Ref.~\onlinecite{aichhorn_quantum_2008}.

\subsection{Variational cluster approach for bosons}
The basic concept of VCA is
that the grand potential $\Omega$ is expressed as a functional of the
self-energy $\mat{\Sigma}$ and that Dyson's equation for the exact
Green's function $\mat{G}$ is recovered at the stationary point of the
self-energy functional
$\Omega[\mat{\Sigma}]$.\cite{potthoff_self-energy-functional_2003,potthoff_self-energy-functional_2003-1}
In order to  evaluate $\Omega[\mat{\Sigma}]$, the self-energy
$\mat{\Sigma}$ of the investigated system is approximated by the
self-energy of an exactly solvable, so-called reference, system. In
practice, 
this means that the self-energy $\mat{\Sigma}$ becomes a
function of the set $\mat{x}$ of single-particle parameters of the
reference system, \ie, $\mat{\Sigma} = \mat{\Sigma}(\mat{x})$. For
bosonic systems the approximated grand potential
reads\cite{koller_variational_2006}
\begin{equation}
 \Omega(\mat{x}) = \Omega^\prime(\mat{x}) + \Tr\,\ln(-\mat{G}^{\prime}(\mat{x})) + \Tr\,\ln(-(\mat{G}_0^{-1}-\mat{\Sigma}(\mat{x}))) \;\mbox{,}
\label{eqn:om}
\end{equation}
where primed quantities  correspond to the reference system and $\mat{G}_0$ is the noninteracting Green's function. The stationary condition on $\Omega (\mat{x})$ is given by
\begin{equation}
 \PDF{\Omega (\mat{x})}{\mat{x}} = 0 \; \mbox{.}
 \label{eq:num:stat}
\end{equation}
This condition can be evaluated numerically by varying some or all of
the single-particle parameters.
In
order to guarantee that a given physical
quantity (such as the number of particles)
is thermodynamically consistent, it is necessary that
$\Omega$ is stationary with respect to the associated coupling
constant (here the chemical
potential).\cite{aichhorn_antiferromagnetic_2006}
Therefore,  varying $\omega_c$ ensures that the total number of photons is
thermodynamically consistent. On the other hand,
 it would be advisable for a conserved
quantity, \ie, $\hat{N}_p$  to be consistent as well.
Otherwise uncommon situations  could occur. For example, as we show
below, the total particle density $\hat{N}_p/N$, evaluated 
as a trace of the Green's functions is not integer in the Mott phase.
This effect becomes stronger close to the tip of the Mott lobe, see \figc{fig:comparison}{b}.
The noninteger particle density, occurring when
$\mu$ is not taken as a variational parameter, clearly introduces an
uncertainty in the determination of the phase boundary.

In principle, however, there is a formal difficulty  in taking
$\mu$ as a variational parameter.
The problem is related to the coupling of $\mu$ with  atomic excitations, which, in contrast to photons, cannot be seen as noninteracting particles.
This is, in general, 
not allowed within VCA, whereby the reference
system can differ from the physical one by a single-particle
Hamiltonian only. The solution is readily overcome by observing that
the two-level atomic system can be mapped onto a hard-core boson
model.
In this way, $\mu$ couples to the total number of ``atomic'' bosons
plus photons, \ie, a noninteracting Hamiltonian.
The hard-core constraint  simply becomes a local (in principle
infinite) interaction, which is
common to the reference and to the physical system.

The mapping of the two-level excitations onto hard-core bosons is mathematically achieved by the following replacements
\begin{alignat*}{2}
 \sigma_i^+ \rightarrow b_i^\dagger \; &\mbox{,} & &\sigma_i^- \rightarrow b_i  \; \mbox{,} \\
 \ket{\downarrow_i} \rightarrow \ket{0_i} \; &\mbox{and} \;\; & &\ket{\uparrow_i} \rightarrow \ket{1_i}  \; \mbox{.}
\end{alignat*}
This is valid provided one excludes states with double occupation of
$b$ particles {\em even as intermediate states}.\cite{hardcore}
With this 
mapping the JCL Hamiltonian reads
\begin{align}
 \hat{H}^{JCL} = &-t \sum_{\left\langle i,\,j \right\rangle} a_i^\dagger \, a_j -\Delta \sum_i b_i^\dagger \, b_i  \nonumber +g\sum_i  ( a_i\,b_i^\dagger  + a_i^\dagger\,b_i ) \\
& - (\mu-\omega_c)\,\hat{N}_p + \lim_{U \rightarrow \infty} \frac{U}{2} \sum_i b_i^\dagger \, b_i (b_i^\dagger \, b_i - 1)  \;\mbox{,}
 \label{eq:spe:jclhc}
\end{align}
where we have formally implemented the hard-core constraint by
introducing an infinite interaction for $b$ particles.
In the restricted Hilbert 
space of zero or one hard-core boson per lattice site,
 the matrix elements of the two representations are identical. 
In principle,
states with higher occupation number $b_i^\dagger \, b_i > 1$ have to
be considered in 
the bosonic version 
as well. However, the occupation of such states would cost infinite 
energy and, therefore, 
they do not influence the energies obtained from the Hilbert space
sector with occupation numbers $b_i^\dagger \, b_i \leq 1$.\cite{hardcore}
We have checked this aspect numerically for very large $U$. It can also
be verified easily when the sector of the Hilbert space with $b_i^\dagger b_i>1$
is included perturbatively. 
These considerations can be straightforwardly extended to
light-matter models with more than one 
atom or atomiclike structure (with two relevant levels)
per cavity. 
In this case, one introduces a boson species for each atom and
the hard-core constraint 
is enforced for each boson species.

In our calculation we take both parameters $\omega_c$ and $\epsilon$ 
of the reference system
as variational parameters ($\mu$ is just a linear combination),
which ensures thermodynamic consistency for the
particle number of both species, 
and, consequently, of the total particle number.
We show below, that varying both parameters instead of just $\omega_c$
provides an 
improvement 
in the accuracy  of the phase boundaries for a given cluster size, see \tab{tab:spe:quality}.

The present formulation of the VCA cannot address the
  superfluid phase, and is, thus, 
restricted to the Mott phase.\cite{koller_variational_2006,
  aichhorn_quantum_2008} 
Outside of the Mott lobes,
peaks with negative (positive) spectral weight appear  in the
positive (negative) $\omega$ region, signaling the instability towards
the superfluid phase.
A treatment of the superfluid phase requires a Nambu Green's
function treatment analogous to the fermionic case.
The boundary of the Mott phase could be, in principle, determined by
this criterion. However, it is simpler (and, of course, equivalent) to 
identify it, for a given $t$, as the region between the ground-state
energies of the 
$(N_p+1)$ and $(N_p-1)$-particle states, which can be directly
inferred from the single-particle spectral function.

In VCA, 
the reference system is chosen to be a decomposition of the total
system into identical clusters, which means that the total lattice of
$N$ sites is divided into clusters of size $L$. Mathematically this
can be described by introducing a superlattice, such that the original
lattice is 
recovered
when a cluster is attached to each lattice site of the superlattice. The reference system defined on a cluster is solved by means of the band Lanczos method. \cite{aichhorn_variational_2006, freund_roland_band_2000} The initial vector of the iterative band Lanczos method for the single-particle excitation term of the cluster Green's function contains $2L$ elements and is given by
\begin{equation}
 \lbrace a_1^\dagger \ket{\psi_0},\,a_2^\dagger \ket{\psi_0} \,\ldots\, a_L^\dagger \ket{\psi_0},\,\sigma_1^+ \ket{\psi_0}\,\ldots\,\sigma_L^+ \ket{\psi_0} \rbrace\;\mbox{,}
 \label{eq:init}
\end{equation}
where $\ket{\psi_0}$ is the $N_p$ particle ground state. For the single-hole excitation term the initial vector of the band Lanczos method is obtained by replacing the creation operators in \eq{eq:init} by annihilation operators.

To evaluate the grand potential and the single-particle Green's function of the original system we use the bosonic $\mat{Q}$-matrix formalism.\cite{knap_spectral_2010} This formalism yields the Green's function $\mat{G}(\tilde{\ve{\ve{k}}},\,\omega)$ in a mixed representation, partly in real space and partly in reciprocal space, see \app~\ref{app:c}. The matrix $\mat{G}(\tilde{\ve{\ve{k}}},\,\omega)$ is of size $2L \times 2L$ and $\tilde{\ve{k}}$ belongs to the first Brillouin zone of the superlattice. Due to the specific order of the creation operators in the initial vector of the band Lanczos method we are able to extract the Green's function for photons $\mat{G}^ {ph}(\tilde{\ve{k}},\,\omega)$ and the Green's function for two-level excitations $\mat{G}^{ex}(\tilde{\ve{k}},\,\omega)$ from $\mat{G}(\tilde{\ve{k}},\,\omega)$ in the following way
\begin{alignat*}{4}
 G_{r,s}^{ph}(\tilde{\ve{k}},\,\omega)&=& &G_{r,s}(\tilde{\ve{k}},\,\omega) &\; &\mbox{and} \\
 G_{r,s}^{ex}(\tilde{\ve{k}},\,\omega)&=& &G_{r+L,s+L}(\tilde{\ve{k}},\,\omega) &\; &\mbox{,}
\end{alignat*}
where $r,\,s \in [1\,\ldots\,L]$.
The application of the periodization prescription proposed in Ref.~\onlinecite{snchal_spectral_2000} (Green's function periodization) yields the fully $\ve{k}$ dependent Green's functions ${G}^{ph}(\ve{k},\,\omega)$ and ${G}^{ex}(\ve{k},\,\omega)$. From that we are able to evaluate the single-particle spectral function
\begin{equation}
 A^x(\ve{k},\,\omega)\equiv-\frac{1}{\pi} \mbox{Im} \, G^x(\ve{k},\,\omega)\;\mbox{,}
 \label{eq:spe:spectralfunction}
\end{equation}
the density of states
\begin{equation}
 N^x(\omega)\equiv \int A^x(\ve{k},\,\omega) \, d\ve{k}  = \frac{1}{N} \sum_{\ve{k}} A^x(\ve{k},\,\omega)
 \label{eq:spe:dos}
\end{equation}
and the momentum distribution
\begin{equation}
 n^x(\ve{k})\equiv - \int_{-\infty}^0 A^x(\ve{k},\,\omega)\,d\omega \;\mbox{,}
\end{equation}
where $x$ can be either $ph$ for photons or $ex$ for two-level excitations. We use the $\mat{Q}$-matrix formalism to evaluate the momentum distribution, since this approach yields particularly accurate results.\cite{knap_spectral_2010} Furthermore we calculate the spatial correlation functions
\begin{equation}
  C^{ph}_{ij} \equiv \langle a_i^\dagger\,a_j \rangle \quad \mbox{and} \quad C^{ex}_{ij} \equiv \langle \sigma_i^+\,\sigma_j^- \rangle \;\mbox{,}
\end{equation}
which just depend on the distance between two cavities $i$ and $j$, \ie, $C^x_{ij} = C^x(|\ve{r}_i-\ve{r}_j|)$.
Notice 
that the poles of the hard-core boson Green's function coincide with
the poles of the two-level excitation Green's function as the energies
of both representations are identical. However, the hard-core boson
Green's function exhibits additional poles located at 
energies of the order $U\to\infty$.
The additional poles which have finite weight result from the fact
that 
excitations such as $b_i^\dagger \, \ket{1_i}$ are
in principal allowed but cost infinite energy,
whereas the corresponding excitation 
$\sigma_i^+ \ket{\uparrow_i}$ is strictly forbidden. 
Therefore,
the single-particle correlation functions $\langle b_{\ve{k}}(t)\,b_{\ve{k}}^\dagger \rangle$ and $ \langle \sigma_{\ve{k}}^-(t)\,\sigma_{\ve{k}}^+ \rangle $ differ
only by contributions from frequencies of the order $U\to\infty$.
Yet it should be mentioned that the single-hole correlations function
of hard-core bosons is not affected by these considerations as
$\langle b_{\ve{k}}^\dagger(t)\,b_{\ve{k}} \rangle$ is always
equivalent to $ \langle \sigma_{\ve{k}}^+(t)\,\sigma_{\ve{k}}^-
\rangle $. This also implies that the spectral weight of the poles
with negative energy are identical for both representations and that
the 
particle density of the two-level system is equal to the particle density of the
hard-core bosons. 
In the following, we will always speak loosely about two-level excitation Green's functions but we have to keep in mind that there are differences in the single-particle spectral weight of the hard-core boson and two-level excitation Green's functions at infinite energies.

\subsection{Polariton properties of the quasiparticles}
In the next step, we want to investigate the polaritonic properties of the JCL model, which arise due to the coupling between the photons and the two-level excitations.

Adding a particle or hole to the many-body ground state may  result in
quasiparticle or collective excitations which are built up by the
$(N_p\pm 1)$-particle eigenstates of the many-body system entering the
Green's function. These many-body eigenstates for the infinite system
can be extracted within the VCA framework from the VCA Green's
function. As shown in \app~\ref{app:c}, they are linear
combinations of the  particle and hole excitations of the cluster
Green's function weighted by the eigenvector matrix $\mat{X}$, defined in \app~\ref{app:c}.

Our goal is to describe the eigenvectors of the ($N_p\pm1$)-particle
Hilbert space, which form the quasiparticle excitations of the 
Green's function
by  polaritonic quasiparticles added to the exact $N_p$ particle groundstate $\ket{\psi_0}$.
To this end, we introduce the polariton creation operators $p_{\alpha,\ve{k}}^\dagger$ for particle excitations and  $h_{\alpha,\ve{k}}^\dagger$  for hole excitations as appropriate linear combinations of photons and two-level excitations
\begin{subequations}
\begin{align}
 p_{\alpha,\ve{k}}^\dagger &= \beta_p^\alpha(\ve{k})\,a_\ve{k}^\dagger + \gamma_p^\alpha(\ve{k})\,\sigma^+_\ve{k} \;\mbox{,} \\
 h_{\alpha,\ve{k}}^\dagger &= \beta_h^\alpha(\ve{k})\,a_\ve{k} + \gamma_h^\alpha(\ve{k})\,\sigma^-_\ve{k} \;\mbox{.}
\end{align}
\label{eq:polCreationOps}
\end{subequations}
It should be stressed that the hole creation operator is not the adjoint of the particle creation operator or its annihilation counterpart, which it would be in the case of noninteracting particles.
As we will see, the coefficients or weights of the linear combinations $\beta_{p/h}^\alpha(\ve{k})$ and $\gamma_{p/h}^\alpha(\ve{k})$ depend on the wave vector $\ve{k}$, the quasiparticle band $\alpha$, and additionally on the filling $n$, which is not explicitly written in \eq{eq:polCreationOps}, since the filling dependence is not important for the present discussions.
The normalized polariton quasiparticle states are defined by
applying the polaritonic operators on the exact $N_p$ particle ground state $\ket{\psi_0}$ yielding
\begin{subequations}\label{eq:xxx1}
\begin{align}
 \ket{\tilde{\psi}_{p,\ve{k}}^\alpha} &= \frac{p_{\alpha,\ve{k}}^\dagger \ket{\psi_0} }{ \sqrt{\bra{\psi_0} p_{\alpha,\ve{k}} \, p_{\alpha,\ve{k}}^\dagger \ket{\psi_0} }} \;\mbox{and} \\
 \ket{\tilde{\psi}_{h,\ve{k}}^\alpha} &= \frac{h_{\alpha,\ve{k}}^\dagger \ket{\psi_0} }{ \sqrt{\bra{\psi_0} h_{\alpha,\ve{k}} \, h^\dagger_{\alpha,\ve{k}} \ket{\psi_0}}} \;\mbox{,}
\end{align}
\label{eq:polwavefu}
\end{subequations}
respectively. The normalization terms can be rewritten as
{\allowdisplaybreaks
\begin{subequations}\label{eq:xxx2}
\begin{align}
 \bra{\psi_0} p_{\alpha,\ve{k}} \, p_{\alpha,\ve{k}}^\dagger \ket{\psi_0} &= {\ve{z}_p^\alpha}^\dagger(\ve{k}) \, \mat{S}_p(\ve{k}) \, \ve{z}_p^\alpha(\ve{k}) \;\mbox{and} \\
 \bra{\psi_0} h_{\alpha,\ve{k}} \, h^\dagger_{\alpha,\ve{k}} \ket{\psi_0} &= {\ve{z}_h^\alpha}^\dagger(\ve{k}) \, \mat{S}_h(\ve{k}) \, \ve{z}_h^\alpha(\ve{k}) \;\mbox{.}
\end{align}
\label{eq:norm}
\end{subequations}
}
In \eq{eq:norm} the vectors $\ve{z}_{p/h}^\alpha(\ve{k})$ are defined as $\ve{z}_{p/h}^\alpha(\ve{k}) \equiv (\beta_{p/h}^\alpha(\ve{k}),\,\gamma_{p/h}^\alpha(\ve{k}))^T$ and $\mat{S}_{p/h}(\ve{k})$ are the overlap matrices of single-particle excitations and single-hole excitations, respectively.
The overlap matrix for the hole excitations is given by
\[ \mat{S}_h(\ve{k})
=\left(\begin{array}{cc} \langle a_\ve{k}^\dagger\,a_\ve{k} \rangle & \langle a_\ve{k}^\dagger\,\sigma^-_\ve{k} \rangle \\
\langle  a_\ve{k}^\dagger\,\sigma^-_\ve{k} \rangle^* &  \langle \sigma^+_\ve{k}\,\sigma^-_\ve{k} \rangle \end{array} \right)
\]
where the static correlation functions  are evaluated in the $N_p$ particle ground state $\ket{\psi_0}$. 
All quantities entering $\mat{S}_h$ are 
correctly evaluated
 in the hard-core boson model
as no excitations of the ``two-level bosons'' into the $n>1$ sector occur.
For the particle case the situation is different, as we need to evaluate
\[
 \mat{S}_p(\ve{k})
 =\left( \begin{array}{cc} \langle a_\ve{k}\,a_\ve{k}^\dagger \rangle & \langle \sigma^-_\ve{k}\,a_\ve{k}^\dagger \rangle^* \\
 \langle \sigma^-_\ve{k}\,a_\ve{k}^\dagger \rangle & \langle \sigma^-_\ve{k} \sigma^+_\ve{k} \rangle \end{array}  \right)\;\mbox{.}
\]
The term $\langle \sigma^-_\ve{k} \sigma^+_\ve{k} \rangle$ of the two-level system 
cannot be directly evaluated
in the hard-core boson model. Using the commutator property $[\sigma^-_i,\sigma^+_j]=0$ for $i\ne j$ and the
local anticommutation relation  
$\{\sigma^-_i,\sigma^+_i\}=1$, we end up with an expression that only contains static correlation functions which can be computed correctly
within the hard-core boson model
\begin{align*}
 \mat{S}_p(\ve{k}) &= \left( \begin{array}{cc}  \langle a_\ve{k}
     a_\ve{k}^\dagger \rangle \quad & \langle
     \sigma^-_\ve{k}\,a_\ve{k}^\dagger \rangle^* \\ \langle
     \sigma^-_\ve{k}\,a_\ve{k}^\dagger \rangle \quad & 1 + \langle \sigma^+_\ve{k}\,\sigma^-_\ve{k} \rangle - \frac{2}{N}\sum_\ve{k} \langle \sigma^+_\ve{k}\,\sigma^-_\ve{k} \rangle \end{array}  \right)\;\mbox{.}
\end{align*}

In order to derive a formalism to construct the optimal polariton weights,
we start out with the analysis of an exact 
eigenvector 
$\ket{\psi^{N_p+1}_{\nu,\ve{k}}}$ of the Hamiltonian in the
$(N_p+1)$-particle sector. For the sake of clarity we will suppress in the following considerations the 
index $\ve{k}$ for all quantities, and the indices $\alpha$ and $p$ for quasiparticle weights and wave functions. The optimality criterion in this case is clearly the overlap of the
exact 
eigenvector
 with the approximate (normalized) vector given in \eqq{eq:xxx1}{eq:xxx2}
\begin{align*}
    \ket{\tilde\psi_{\nu}} = \frac{1}{\sqrt{{\ve z^\nu}^\dagger \, \mat{S}_p \, \ve z^\nu}}
    \sum_I z_I^\nu \, d_I^\dagger \ket{\psi_0}\;,
\end{align*}
where $I$ denote the components of the two-dimensional vectors, and
$d_{1,\ve{k}}\equiv a_\ve{k}$ and $d_{2,\ve{k}}\equiv \sigma_\ve{k}^-$, see \app~\ref{app:c}.
The  maximization of $|\braket{\psi^{N+1}_\nu|\tilde\psi_\nu}|^2$ leads to the generalized eigenvalue problem
\begin{align}\label{eq:gen_ev}
    \mat{A}^\nu \, \tilde{\ve z}^\nu &= \lambda \, \mat{S}_p \, \tilde{\ve z}^\nu\;\mbox{,}
\end{align}
where the elements of the $2 \times 2$ matrix $\mat{A}^\nu$ are
\[
A_{IJ}^\nu= \bra{\psi_0} d_{I} \ket{\psi^{N_p+1}_\nu}
    \bra{\psi^{N_p+1}_\nu} d_{J}^\dagger \ket{\psi_0}\;\mbox{.}\]
In \eq{eq:gen_ev} we replaced $\ve{z}^\nu$ by $\tilde{\ve{z}}^\nu$ as the eigenvalues are just determined except for a constant $Z$, which will be specified later.
As the eigenvalue corresponds to the value of the overlap squared
$\lambda = |\braket{\psi^{N+1}_\nu|\tilde\psi_\nu}|^2$, 
the 
deviation 
of the eigenvalue 
from 
one is a measure of the quality of the polariton approximation. It also points out that the eigenvector corresponding to the largest eigenvalue determines the optimal polariton coefficients.
Interestingly, $A^\nu_{IJ}$ is the contribution of the excitation
$\nu$ to the corresponding spectral function, 
\ie, its {\em quasiparticle weight}.
In general, the
quasiparticle peak is  a superposition of several exact many-body
eigenstates. Hence, the obvious generalization of the optimality
criterion is to sum over all eigenstates $\nu$, which contribute to
the  quasiparticle excitation $\alpha$. 
To this end we define an energy window $\Omega_\alpha$ in which the quasiparticle peak $\alpha$ is located and we integrate  the spectral density
in this energy window resulting in
\begin{align*}
   \tilde A_{IJ}(\ve{k},\,\Omega_\alpha) \equiv \sum_ {\nu,\,\omega_\nu(\ve{k}) \in \Omega_\alpha}
    A^\nu_{IJ}\;.
\end{align*}
The polariton coefficients are again obtained by the generalized eigenvalue problem
\begin{align*}
    \tilde{\mat{A}}(\ve{k},\,\Omega_\alpha) \, \tilde{ \ve z} = \lambda \, \mat{S}_p \, \tilde{\ve z}\;\mbox{.}
\end{align*}
and the eigenvalue is given by
\begin{align}\label{eq:lambda}
    \lambda &= \frac{\tilde{\ve z}^\dagger \, \tilde{\mat{A}}(\ve{k},\,\Omega_\alpha) \, \tilde{ \ve z}}{\tilde{ \ve z}^\dagger \, \mat{S}_{p} \, \tilde{ \ve z}}\;\mbox{.}
\end{align}
The eigenvalues are still restricted to the unit interval $[0,1]$. The lower limit is due to the positivity of $\tilde{\mat{A}}$ and $\mat{S}_p$. The upper limit follows from the property that
a summation of the integrated spectral density over all  nonoverlapping energy intervals $\Omega_\alpha$ is given by
\begin{align*}
    \sum_\alpha^\text{particles}   \tilde  A_{IJ}(\ve{k},\Omega_\alpha) &= \braket{d_{I,\ve{k}} d^\dagger_{J,\ve{k}}} = (\mat{S}_p)_{IJ}\,\mbox{.}
\end{align*}
Of course, $\tilde{\ve z}$ and, hence, the polariton operators will depend on the wave vector $\ve{k}$, the quasiparticle band index $\alpha$ and the filling $n$, \ie, the Mott lobe.
The discussion so far was for the particle case only, however, it is 
straightforward to iterate the procedure for the hole case.

Eventually, we merely need the integrated spectral density
$\mat{A}(\ve{k},\,\Omega_\alpha)$ determined within the VCA 
framework,
 which is given by
\begin{align*}
    \tilde A_{IJ}(\ve{k},\Omega_\alpha)
    &\equiv- \sum_{\nu, \, \omega_\nu(\ve{k})\in \Omega_\alpha}
    (\tilde{\mat{Q}} \mat{X})_{I,\nu}
        (\mat{X}^{-1} \mat{S} \tilde{\mat{Q}}^\dagger)_{\nu,J}\;.
\end{align*}
Details are presented in \app~\ref{app:c} as well as the proof that all contributions of the sum have the same sign, which is necessary for the optimality criterion to make sense at all.
The optimality criterion as well as the eigenvalue problem only fix the coefficient vector $\ve z$ up to a normalization factor $Z$, \ie, $\ve z = Z\,\tilde{\ve z}$.
The latter is determined by the condition that the total spectral weight should be conserved
\begin{equation}
 Z^2\,\tilde{\ve{z}}^\dagger \, \tilde{\mat{A}} \, \tilde{\ve{z}} \stackrel{!}{=} \tr \,\tilde{\mat{A}} \;\mbox{.}
\label{eq:trconv}
\end{equation}
As the excitations can now be described by 
wave vector, band and filling dependent 
polaritonic quasiparticles, it remains to evaluate the polariton
spectral function $A^p(\ve{k},\,\omega)$, which is 
due to the invariance of the trace
in \eq{eq:trconv} equal to the sum of the photon spectral function $A^{ph}(\ve{k},\,\omega)$ and the two-level excitation spectral function $A^{ex}(\ve{k},\,\omega)$.

\section{\label{sec:results}Results}
In this section, we present the results of our calculations. Specifically,
in Sec.~\ref{qpt}, we discuss the quantum phase transition from Mott
  phase to superfluid phase occurring in the JCL model and investigate
  the impact of the variational parameter space on the accuracy of the
  results. 
In Sec.~\ref{spec}, we study the spectral properties of both photons
  as well as two-level excitations. The first two subsections refer to
  results obtained for zero detuning $\Delta=0$, whereas nonzero
  detuning is considered in the third subsection, Sec.~\ref{nonz}. Finally, in Sec.~\ref{pola}, 
we study the
  polaritonic properties of the JCL model. In particular, we introduce
  polariton quasiparticles as wave vector and filling dependent linear
  combinations of photons and two-level excitations and analyze the
  weights of their constituents.

\subsection{Quantum phase transition}
\label{qpt}
\begin{figure}
        \centering
        \includegraphics[width=0.48\textwidth]{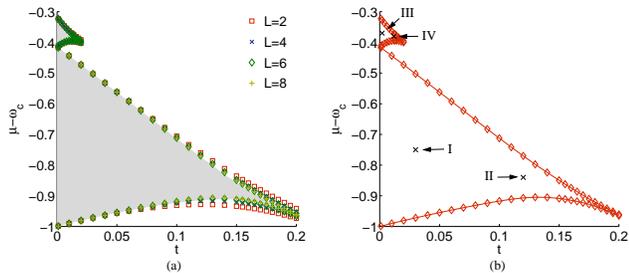}
        \caption{(Color online) Phase boundaries of the JCL model in one dimension for zero detuning $\Delta=0$. \fc{a} VCA results for the variational parameters $\mat{x}=\lbrace \omega_c,\, \epsilon \rbrace$ and various cluster sizes of the reference system. The gray shaded area indicates DMRG data.\cite{rossini_mott-insulating_2007} \fc{b} Phase boundaries obtained for the largest cluster ($L=8$ for the first Mott lobe and $L=6$ for the second Mott lobe). The marks refer to parameters where spectral functions are evaluated.  }
        \label{fig:pd}
\end{figure}

\begin{figure}
        \centering
        \includegraphics[width=0.48\textwidth]{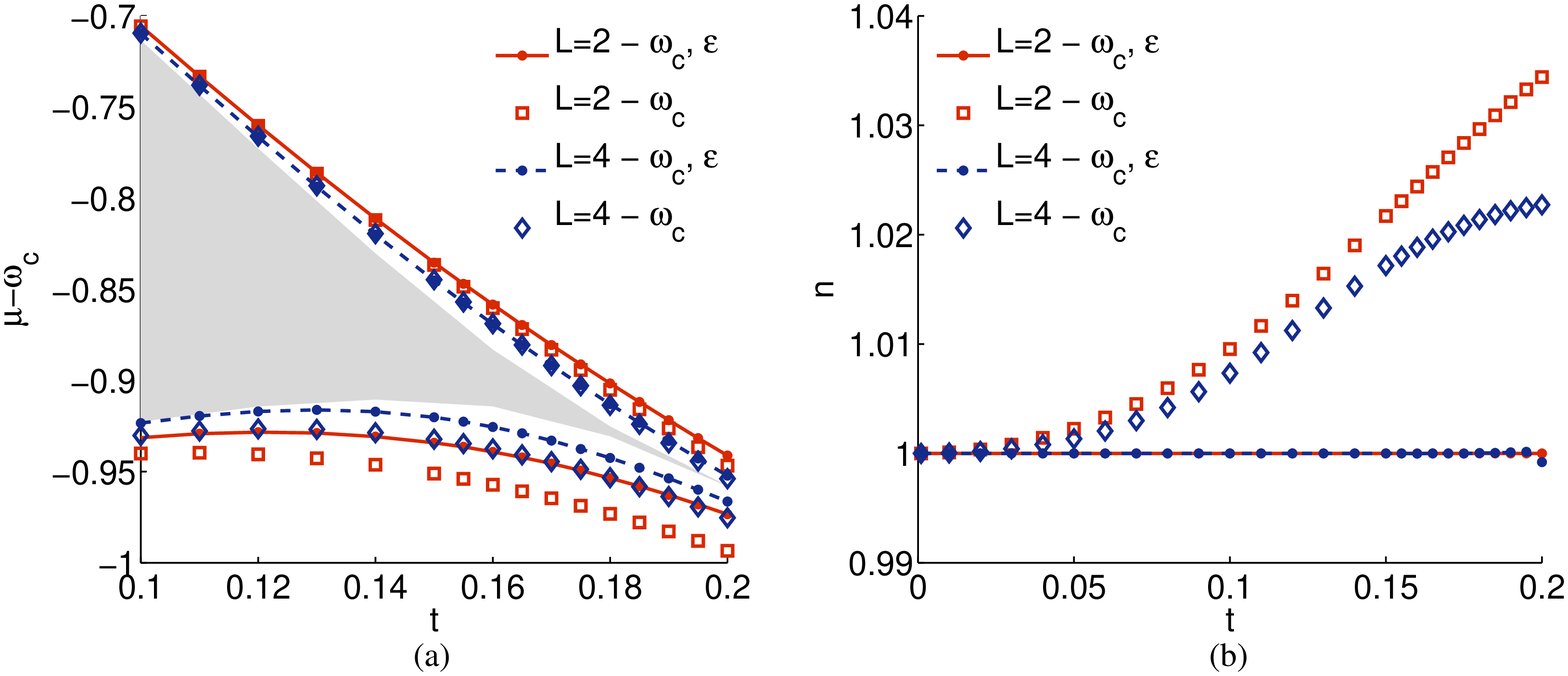}
        \caption{(Color online) Comparison between the results obtained with the variational parameters $\mat{x}=\lbrace \omega_c,\, \epsilon \rbrace$ and  $\mat{x}=\lbrace \omega_c \rbrace$, respectively, for small clusters of size $L=2$ and $L=4$. \fc{a} Phase boundaries at the tip of the first Mott lobe. The gray shaded area indicates DMRG results.\cite{rossini_mott-insulating_2007} \fc{b}~Total particle density $n$, which is the sum of the photon density and the two-level excitation density, across the first Mott lobe. }
        \label{fig:comparison}
\end{figure}

The JCL model exhibits, comparable to the BH model,\cite{fisher_boson_1989} a quantum phase transition from a localized Mott phase to a delocalized superfluid phase.
For integer particle density and small hopping 
strength $t$,
 the ground state of the system is a Mott state. The first two Mott lobes of the one-dimensional (1D) JCL model for zero detuning $\Delta=0$ obtained by means of VCA with the variational parameters $\mat{x}=\lbrace \omega_c,\, \epsilon \rbrace$ are shown in \fig{fig:pd}.
%fig:pd
As discussed in the previous section, including $\epsilon$ in the set
of variational parameters is nontrivial and is solely possible since
the two-level excitations can be mapped onto hard-core bosons. The
gray shaded area in \figc{fig:pd}{a} indicates DMRG results for the
phase boundary obtained by D.~Rossini \textit{et~al.} in
Ref.~\onlinecite{rossini_mott-insulating_2007}. We find excellent
agreement between the phase boundary evaluated by means of VCA with
the variational parameter set $\mat{x}=\lbrace \omega_c,\, \epsilon
\rbrace$ and the DMRG results, even at the lobe tip, where 
quantum fluctuation
effects are most important, and even for moderate cluster sizes $L \gtrsim 4$.
\Figc{fig:comparison}{a} compares the phase boundaries at the tip of the first Mott lobe for different variational parameters.
%fig:comparison
The results obtained with $\mat{x}=\lbrace \omega_c,\, \epsilon
\rbrace$ are connected by lines, whereas the open symbols correspond
to $\mat{x}=\lbrace \omega_c \rbrace$. We observe that using both
on-site energies as variational parameters 
improves the results for the phase boundary and also yields a better approximation for the slope of the lobe tip. A quantitative measure for the quality $\chi$ of the calculated phase boundary is given by the absolute deviation from the DMRG data per phase boundary point
\begin{equation}
 \chi = \frac{1}{M_p} \sum_i \left| p_i^{V} - p_i^{D} \right| \,\mbox{,}
 \label{eq:spe:chiQuality}
\end{equation}
where $p_i^{V}$ and $p_i^{D}$ are corresponding phase boundary points calculated by means of VCA and DMRG, respectively, and $M_p$ is the number of phase boundary points, which contribute to the sum. In \tab{tab:spe:quality} we compare the quality $\chi/10^{-3}$ of the phase boundary between the two sets of variational parameters for various cluster sizes.
\begin{table}
        \caption{ Quality $\chi/10^{-3}$ of the phase boundary for $\mat{x}=\lbrace \omega_c \rbrace $ and $\mat{x}=\lbrace \omega_c,\,\epsilon \rbrace $, respectively. The quality $\chi$ is evaluated using \eq{eq:spe:chiQuality}. }
        \label{tab:spe:quality}
        \centering
        \begin{tabular}{rcl}
    $L$ & \ldots & number of cluster sites  \\
    $\epsilon,\, \omega_c$ & \ldots & variational parameters \\
    \emph{IMP} & \ldots &  improvement in quality when using the variation-\\ &&al parameters $\mat{x}=\lbrace \omega_c,\,\epsilon \rbrace $ instead of $\mat{x}=\lbrace \omega_c \rbrace $ \\

        \end{tabular} \newline
        \begin{tabular}{|c||r @ {.} l|r @ {.} l|c|}
                \hline
                \emph{$L$} & \multicolumn{2}{c|}{\emph{$\lbrace \omega_c\rbrace $}} & \multicolumn{2}{c|}{\emph{$\lbrace \omega_c,\, \epsilon\rbrace $}} & \emph{IMP} \\
                \hline
                \hline
                2 & 15&95  & {\;\;}11&34 & 1.41 \\ \hline
                4 & 8&20 & 4&92 & 1.67 \\ \hline
                6 & 5&34 & 3&16 & 1.69 \\ \hline
                8 & 3&95 & 3&07 & 1.29 \\ \hline
        \end{tabular}
\end{table}
When using the augmented set of variational parameters $\mat{x}=\lbrace \omega_c,\,\epsilon \rbrace $ in contrast to $\mat{x}=\lbrace \omega_c \rbrace $ we observe an improvement in the quality of the phase boundary which ranges from $1.3$ to $1.7$ depending on the cluster size of the reference system.
Using both the resonance frequency $\omega_c$ of the cavities and the
energy spacing $\epsilon$ of the two-level system as variational parameters thus
provides a significant improvement 
with respect to 
the case of a single variational parameter.\cite{aichhorn_quantum_2008}
As discussed in Refs.~\onlinecite{koller_variational_2006} and
\onlinecite{aichhorn_antiferromagnetic_2006},
a correct particle
density in the original system can only be obtained when 
the corresponding
on-site
energies are included in the set of variational parameters, \ie,
in the case of the JCL model
$\mat{x}=\lbrace \omega_c,\, \epsilon \rbrace$. This is demonstrated
in \figc{fig:comparison}{b}, where the total particle density $n$,
which consists of a photon and a two-level excitation contribution,
is evaluated along the first Mott lobe. For $\mat{x}=\lbrace \omega_c
\rbrace$ the deviation of the particle density from one is growing
with increasing hopping strength $t$ but shrinking with increasing
cluster size $L$. However, when $\epsilon$ is included as variational
parameter the total particle density $n$ is as desired equal to one
across the whole first Mott lobe. A deviation 
of about $0.001$ can be
observed for $t=0.2$. Yet, the hopping strength $t=0.2$ is probably
even slightly above the critical hopping strength $t^*$, which
indicates the tip of the Mott
lobe.\cite{rossini_mott-insulating_2007} 

The phase diagram of the 1D JCL model is in many aspects similar to the phase diagram of the 1D BH model.\cite{koller_variational_2006, khner_one-dimensional_2000} Particularly, the Mott lobes are point shaped and a reentrance behavior can be observed, which means that for certain values of $\mu$ upon increasing $t$ the system leaves the Mott phase and later on enters it again. Yet a very important difference is that the width of the lobes of the JCL model at zero hopping is shrinking with increasing particle density. This comes from the fact that the effective on-site repulsion of the JCL lattice model, which is hidden in the interaction between photons and two-level excitations, is not constant, as in the Bose-Hubbard model. The exact location of the phase boundaries at zero hopping is derived as a by-product in \app~\ref{app:a}, whose major intention is, however, to introduce the notation used for the dressed states $\ket{n,\,\alpha}$ and for the corresponding energies $E_{\ket{n,\,\alpha}}$, where $\alpha \in \lbrace -,\,+ \rbrace$ describing the ground state and the excited state in the corresponding constant particle number sector of the single-cavity Hilbert space.

\subsection{Spectral properties of photons and two-level excitations}
\label{spec}
\begin{figure}
        \centering
        \includegraphics[width=0.48\textwidth]{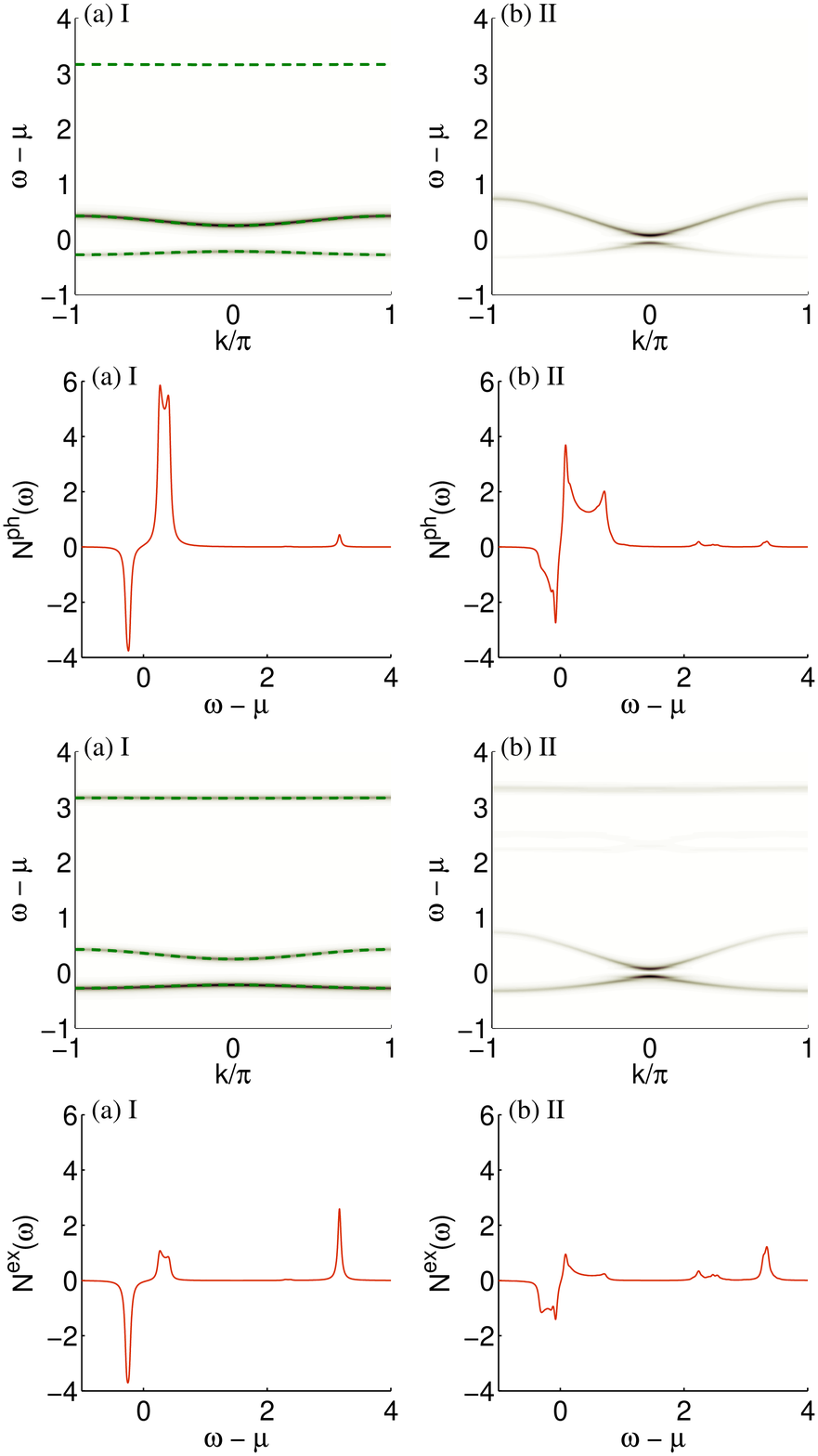}
        \caption{(Color online) Photon spectral function $A^{ph}(\ve{k},\,\omega)$, first row, and density of states $N^{ph}(\omega)$, second row. Two-level excitation spectral function $A^{ex}(\ve{k},\,\omega)$, third row, and density of states $N^{ex}(\omega)$, fourth row. The spectral functions are evaluated for the parameters \fc{a} $t=0.03$, $\mu-\omega_c=-0.75$, $\Delta=0$ and \fc {b} $t=0.12$, $\mu-\omega_c=-0.84$, $\Delta=0$, which belong to the first Mott lobe. The dashed lines in the spectral functions in \fc{a} correspond to first-order degenerate perturbation theory results, see \app~\ref{app:b}. The Roman numerals in the captions of the subfigures refer to the marks in \figc{fig:pd}{b}.}
        \label{fig:sfLobe1}
\end{figure}

The spectral function for photons $A^{ph}(\ve{k},\,\omega)$, the spectral function for two-level excitations $A^{ex}(\ve{k},\,\omega)$ and the corresponding densities of states $N^{ph}(\omega)$ and $N^{ex}(\omega)$ evaluated by means of VCA for parameters belonging to the first Mott lobe are shown in \fig{fig:sfLobe1}.
%fig:sfLobe1
We use an artificial broadening $\eta = 0.03$ and the variational parameter set $\mat{x}=\lbrace \omega_c,\,\epsilon \rbrace $ for the numerical evaluation of the spectral functions.
Both spectral functions $A^{ph}(\ve{k},\,\omega)$ and $A^{ex}(\ve{k},\,\omega)$ have the same gap as the photons and the two-level excitations are coupled.
The spectral functions of the JCL model generally consist of four bands.
This can best  be understood in terms of the analytic solution of the JCL model for zero hopping strength $t=0$.
The ground state $\ket{\psi_0}$ of the JCL model in the Mott phase with particle density $n$ for zero hopping is given by the tensor product state
\begin{equation}
 \ket{\Psi_0} = \bigotimes_{\nu=1}^N \ket{n,\,-}_\nu \; \mbox{,}
\end{equation}
where $\ket{n,\,-}_\nu$ is the dressed $n$ particle ground state of lattice site $\nu$.
The states with a single-particle excitation are those, where $N-1$ sites remain in the dressed state $\ket{n,\,-}$ and one site is excited to the state $\ket{n+1,\,\alpha}$. Similarly, for the single-hole excitation $N-1$ sites remain in the state $\ket{n,\,-}$ and one site is excited to the state $\ket{n-1,\,\alpha}$. In both cases, the excited states are $N$ fold degenerate as the particle/hole excitation can be located on any of the $N$ lattice sites. The degenerate states have thus the structure
{\allowdisplaybreaks
\begin{subequations}
\begin{align}
 \ket{\Psi_p^{\alpha,\,l}} &\equiv \ket{n+1,\,\alpha}_l \bigotimes_{ \genfrac{}{}{0pt}{}{\nu = 1}{\nu \neq l} }^N \ket{n,\,-}_\nu \;  \mbox{and} \label{eq:psi1} \\
 \ket{\Psi_h^{\alpha,\,l}} &\equiv \ket{n-1,\,\alpha}_l \bigotimes_{ \genfrac{}{}{0pt}{}{\nu = 1}{\nu \neq l} }^N \ket{n,\,-}_\nu \; \mbox{,} \label{eq:psi2}
\end{align}
\label{eq:psiDegenerate}
\end{subequations}
}
respectively. Two of the four bands, we refer to them as lower modes $\omega_{p/h}^-$, emerge from the excitation of site $i$ from the dressed state $\ket{n,\,-}_i$ to the states $\ket{n\pm1,\,-}_i$, which are ground states of the corresponding Hilbert-space sector with constant particle number. Analogously, we refer to the bands which emerge from the excitation of site $i$ from $\ket{n,\,-}_i$ to the excited states in the corresponding particle sector $\ket{n\pm1,\,+}_i$ as upper modes $\omega_{p/h}^+$. The presence of the upper modes has been first noted by S.~Schmidt \textit{et al.} in Ref.~\onlinecite{schmidt_strong_2009} and has been numerically observed in latest QMC calculations\cite{pippan_excitation_2009} as well. The two upper modes $\omega_{p/h}^+$ indicate a clear deviation from the BH physics, which emerges due to the composition of two distinct particles. As discussed in the previous section, the two particle bands $\omega_p^\alpha$, $\alpha \in \lbrace -,\,+ \rbrace $, determine the polariton particle creation operators $p_{\alpha,\ve{k}}^\dagger$ whereas the two hole bands $\omega_h^\alpha$ specify the hole creation operators $h_{\alpha,\ve{k}}^\dagger$.

In the spectral functions of \fig{fig:sfLobe1}, the lower modes $\omega_{p/h}^-$ correspond to the cosinelike shaped bands centered around $\omega-\mu = 0$. The intensities of the lower modes $\omega_{p/h}^-$ are contrary for the photon spectral function $A^{ph}(\ve{k}, \,\omega)$ and the two-level excitation spectral function $A^{ex}(\ve{k}, \omega)$. For $A^{ph}(\ve{k}, \,\omega)$ the particle band $\omega_{p}^-$ is more intense than the hole band $\omega_{h}^-$ whereas the hole band is more intense than the particle band for $A^{ex}(\ve{k}, \omega)$. For the first Mott lobe the upper hole mode $\omega_{h}^+$ does not exist as this would require to excite a single-site $i$ from the dressed state $\ket{1,\,-}_i$ to the non-existing state $\ket{0,\,+}_i$. Thus, only the upper particle mode $\omega_{p}^+$ can be observed in the spectral functions shown in \fig{fig:sfLobe1}, which corresponds to the essentially flat band located at $\omega - \mu \approx 3$. In \app~\ref{app:b}, we evaluate the single-particle and single-hole excitation bands by means of first-order degenerate perturbation theory, which yields
\begin{subequations}
\begin{align}
 \omega_{p,1}^\alpha &= (\omega_c - \mu)+ \alpha\,q(n+1)+q(n) - 2 \, \tilde{t}_p^\alpha \cos k \;\mbox{and} \label{eq:omParticle} \\
 \omega_{h,1}^\alpha &= (\omega_c - \mu)-\alpha\,q(n-1)-q(n) + 2 \, \tilde{t}_h^\alpha \cos k \;\mbox{,} \label{eq:omHole}
\end{align}
\label{eq:om}
\end{subequations}
respectively, where  $\tilde{t}_{p/h}^\alpha$ is the renormalized hopping strength. \Figc{fig:sfLobe1}{a} shows, additionally to the spectral functions obtained by means of VCA, the perturbation results for the bands. For small hopping strength we observe, as expected, good agreement between the two approaches. From the analytic solution of the bands we are able to extract their width, which is given by $2\,\tilde{t}_{p/h}^\alpha$. The renormalization factor in $\tilde{t}_{p/h}^\alpha$ essentially consists of a square of the form $(a+b)^2$, see \eqq{eq:teffp}{eq:teffh}. Evaluating these expressions shows that $a,\,b > 0$ for the lower modes $\omega_{p/h}^-$ but $a>0$ and $b<0$ for the upper modes $\omega_{p/h}^+$. Therefore, $a$ and $b$ almost cancel each other in the latter case, which yields a small renormalized hopping strength of the upper modes $\tilde{t}_{p/h}^+$ in comparison to the one of the lower modes $\tilde{t}_{p/h}^-$ and thus, essentially flat upper particle/hole bands $\omega_{p/h}^+$.\cite{schmidt_strong_2009}
Plugging in the value of the modified chemical potential $\mu-\omega_c
= -0.75$, which has been used to evaluate the spectral function shown
in \figc{fig:sfLobe1}{a}, into \eq{eq:omParticle} yields $
\omega_{p,1}^+ \approx 3.16 $, where we neglected the dependence on
the wave vector. This matches perfectly with the VCA results. In
  addition to previous work\cite{schmidt_strong_2009,
    pippan_excitation_2009} we evaluate the upper modes not only for
  photons but also for two level-excitations. Interestingly, the
  spectral weight differs significantly for the two types of particles. In
  particular, the upper particle mode $\omega_p^+$ has a very large
intensity in the two-level excitation spectral function $
A^{ex}(\ve{k},\,\omega)$, but is almost not visible in the photon
spectral function $A^{ph}(\ve{k},\,\omega)$. 
\begin{figure}
        \centering
        \includegraphics[width=0.48\textwidth]{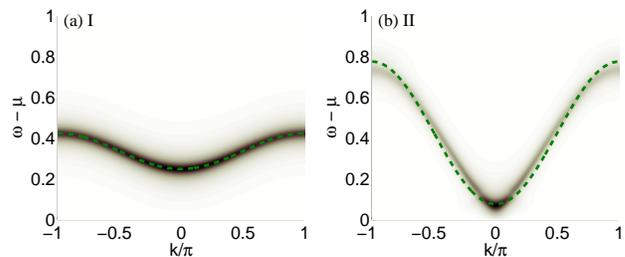}
        \caption{(Color online) Extract of the lower particle band $\omega_p^-$ of the photon spectral functions $A^{ph}(\ve{k},\,\omega)$ shown in \fig{fig:sfLobe1}, where the parameters \fc{a} $t=0.03$, $\mu-\omega_c=-0.75$, $\Delta=0$ and \fc {b} $t=0.12$, $\mu-\omega_c=-0.84$, and $\Delta=0$ have been used. VCA results (density plot) are compared with bands evaluated by means of first-order degenerate perturbation theory (dashed lines).  }
        \label{fig:sfPert}
\end{figure}
For the spectral function shown in \figc{fig:sfLobe1}{b} a different chemical potential $\mu-\omega_c=-0.84$ has been used. Thus, the upper particle mode is shifted slightly upwards in comparison to \figc{fig:sfLobe1}{a} and is located at $ \omega_{p,1}^+ \approx 3.25 $.
\Fig{fig:sfPert} shows the lower particle band $\omega_p^-$ of the photon spectral function $A^{ph}(\ve{k},\,\omega)$ for the same parameters as in \fig{fig:sfLobe1}. In this figure we compare the VCA results for different hopping strengths with the results obtained by means of first-order degenerate perturbation theory. For small hopping strength, $t=0.03$, see \figc{fig:sfPert}{a}, the perturbative results agree very well with the VCA results in both the width as well as the shape of the band. However, for large hopping strength $t=0.12$, which is already close to the tip of the Mott lobe, the lower particle band does not exhibit a simple cosine shape anymore, see \figc{fig:sfPert}{b}. In addition the width of the band is slightly overestimated by first-order degenerate perturbation theory.
%fig:sfPert

\begin{figure}
        \centering
        \includegraphics[width=0.48\textwidth]{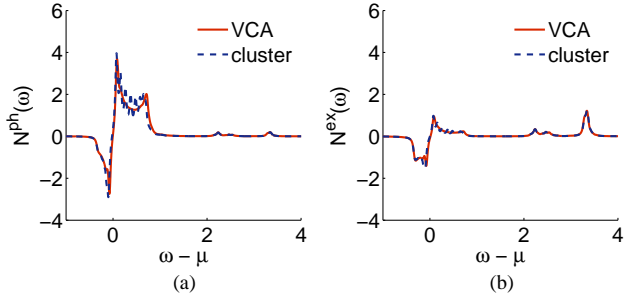}
        \caption{(Color online) Comparison between the density of states obtained from the VCA Green's function, solid lines, and the density of states obtained from the cluster Green's function, dashed lines. \fc{a} density of states of photons $N^{ph}(\omega)$ and \fc{b} density of states of two-level excitations $N^{ex}(\omega)$. The parameters used for these plots are the same as in \figc{fig:sfLobe1}{b}. }
        \label{fig:dosCluster}
\end{figure}

In the spectral functions shown in \figc{fig:sfLobe1}{b} there is additional spectral weight located at $\omega-\mu \approx 2$. We can exclude that this additional weight stems from the periodization prescription used in VCA or from any other VCA internal processes as it also appears in the cluster Green's function, which is solved by exact diagonalization. This can be verified best by comparing the density of states obtained from the VCA Green's function with the density of states obtained from the cluster Green's function, see \fig{fig:dosCluster}.
%fig:dosCluster
\begin{figure}
        \centering
        \includegraphics[width=0.48\textwidth]{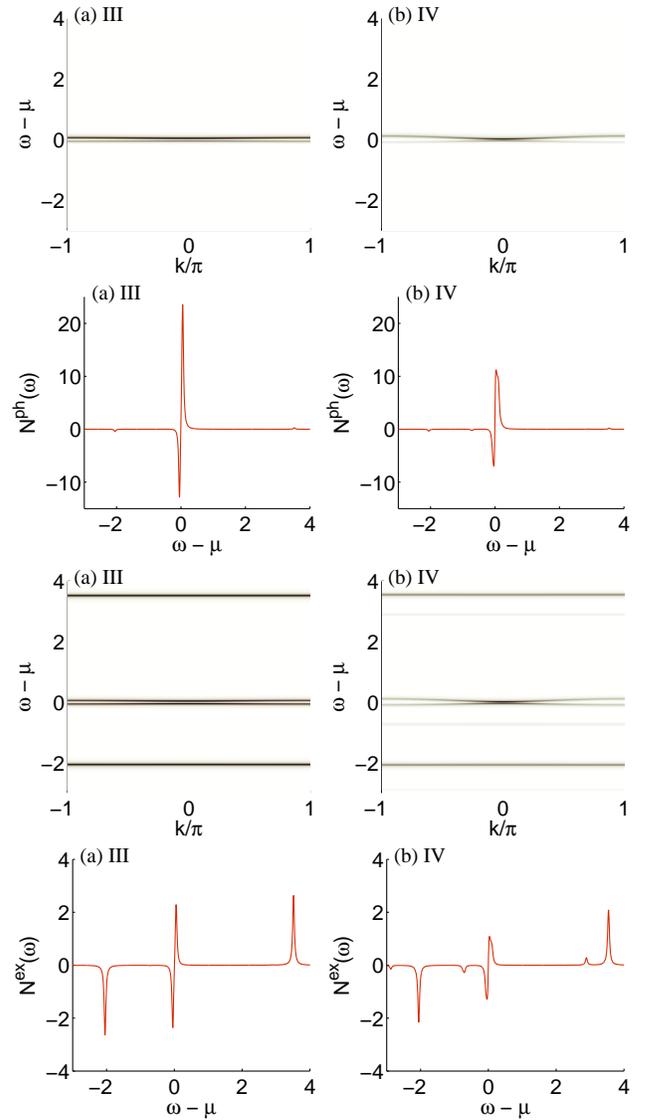}
        \caption{(Color online) Photon spectral function $A^{ph}(\ve{k},\,\omega)$, first row, and density of states $N^{ph}(\omega)$, second row. Two-level excitation spectral function $A^{ex}(\ve{k},\,\omega)$, third row, and density of states $N^{ex}(\omega)$, fourth row. The spectral functions are evaluated for the parameters \fc{a} $t=0.002$, $\mu-\omega_c=-0.37$, $\Delta=0$ and \fc {b} $t=0.012$, $\mu-\omega_c=-0.38$, and $\Delta=0$, which belong to the second Mott lobe. The Roman numerals in the captions of the subfigures refer to the marks in \figc{fig:pd}{b}. }
        \label{fig:sfLobe2}
\end{figure}
Both densities of states, the one obtained from the cluster Green's function and the one obtained from the VCA Green's function, exhibit a peak located at $ \omega-\mu \approx 2 $.
The additional peak can be revealed in the 
framework 
of perturbation theory. 
First-order 
local particle fluctuations in the ground state will have contributions of the form
\begin{align*}
\ket{\Delta \psi^{(1)}} &= \frac{t}{\Delta E}  \ket{n+1,\,\alpha}_l\otimes \ket{n-1,\,\beta}_{l'} \bigotimes_{ \genfrac{}{}{0pt}{}{\nu = 1}{\nu \neq l,l'} }^N \ket{n,\,-}_\nu \;,
\end{align*}
where $l,l'$ correspond to nearest-neighbor sites.
Due to the energy denominator $\Delta E$ the predominant terms are those with
$\alpha=\beta=-$. The correction term $\ket{\Delta \psi^{(1)}}$ is proportional to the hopping strength $t$, which explains, why the additional peak is not present in
\figc{fig:sfLobe1}{a}. The particle excitation couples to final states with an additional particle either on site $l$, $l'$ or on one of the remaining sites.
A detailed analysis shows that the excitation, responsible for the additional peak
at about $\omega-\mu \approx 2$, is
\begin{align*}
\ket{\psi^{N_p+1}} =  \ket{n+1,\,-}_l\otimes \ket{n,\,+}_{l'} \bigotimes_{ \genfrac{}{}{0pt}{}{\nu = 1}{\nu \neq l,l'} }^N \ket{n,\,-}_\nu \;.
\end{align*}
The corresponding excitation energy is given by
{\allowdisplaybreaks
\begin{align*}
    \tilde{\omega}_p &= E_{\ket{n+1,-}}+ E_{\ket{n,+}} + (N-2)E_{\ket{n,-}} - E^{N}_0\\
    &= E_{\ket{n+1,-}}+  E_{\ket{n,+}} -2E_{\ket{n,-}}\\
    &= \omega_c-\mu - q(n+1) + 3q(n) \;.
\end{align*}
}
For zero detuning and $\mu-\omega_c=-0.84$ the energy is $\tilde{\omega}_p = \omega-\mu = 2.4$.

As discussed before the upper hole mode $\omega_{h}^+$ does not exist in the first Mott lobe. Yet, the mode $\omega_h^+$ is present in spectral functions of the second Mott lobe, see \fig{fig:sfLobe2}.
%fig:sfLobe2
According to \eq{eq:om} the upper modes are located at $\omega_{p,1}^+ \approx 3.52$ and $\omega_{h,1}^+ \approx -2.04$ for the parameters used in \figc{fig:sfLobe2}{a}. This matches very well the results obtained by means of VCA. The chemical potential of the spectral function shown in \figc{fig:sfLobe2}{b} differs from the one of \fc{a} merely about 0.01. Thus, the bands $\omega_{p/h}^+$ are located at rather the same position in both spectral functions.

\begin{figure}
        \centering
        \includegraphics[width=0.48\textwidth]{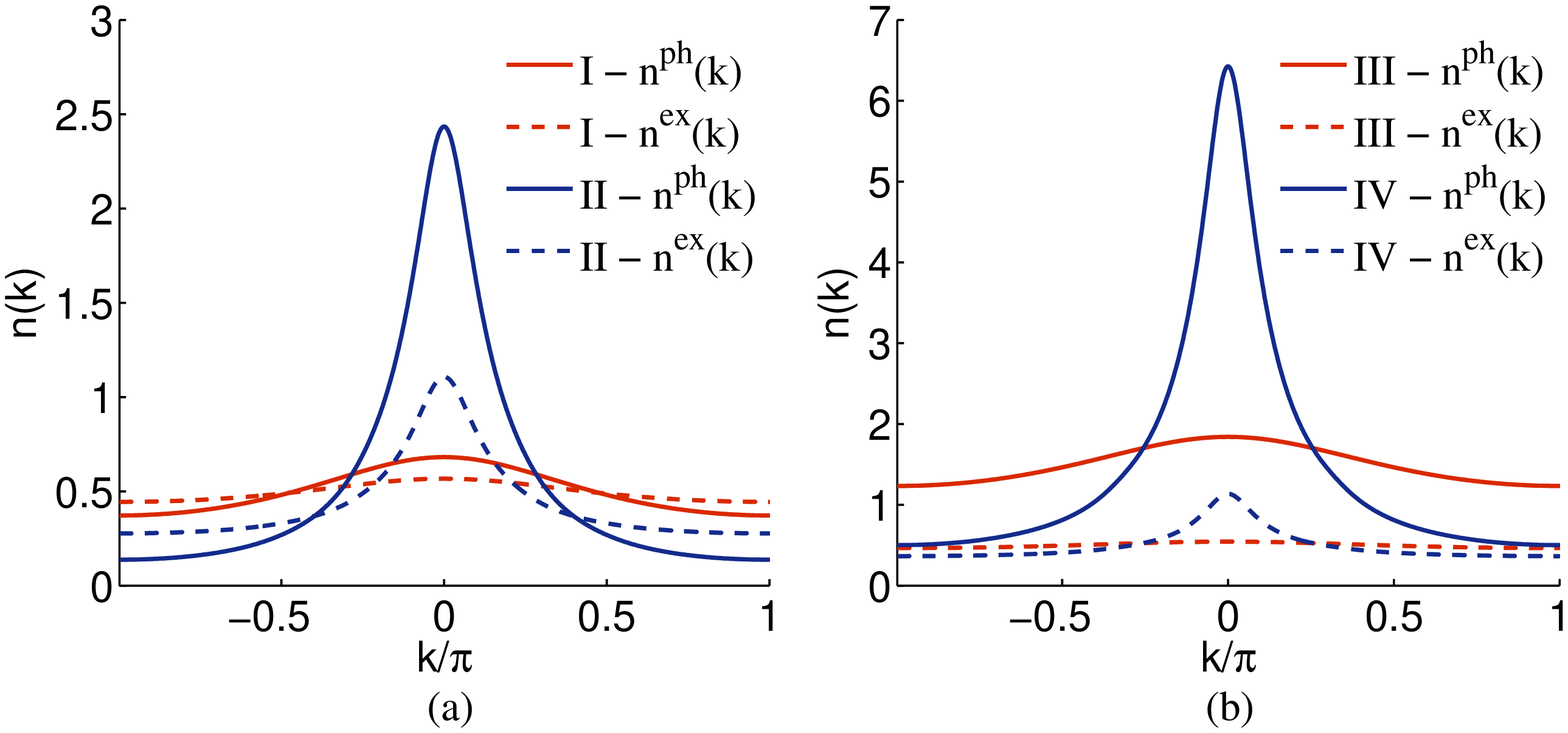}
        \caption{(Color online) Momentum distribution \fc{a} in the first Mott lobe and \fc{b} in the second Mott lobe for the parameters marked with Roman numerals in \figc{fig:pd}{b}. Solid lines correspond to the momentum distributions of photons $n^{ph}(\ve{k})$ and dashed lines to the momentum distributions of two-level excitations $n^{ex}(\ve{k})$. }
        \label{fig:n}
\end{figure}

The momentum distribution for photons $n^{ph}(\ve{k})$ and two-level excitations $n^{ex}(\ve{k})$ in the first and second Mott lobe are shown in \fig{fig:n}.
%fig:n
For increasing hopping strength $t$ the momentum distribution becomes
more peaked for both the photons and the two-level excitations. In the
first Mott lobe the momentum distributions $n^{ph}(\ve{k})$ and
$n^{ex}(\ve{k})$ are centered around $0.5$, which means that the
cavities are on average equally occupied by photons and two-level
excitations. In the second Mott lobe $n^{ph}(\ve{k})$ is centered
around $1.5$. However, 
$n^{ex}(\ve{k})$ is still centered around $0.5$, as the maximum local
occupation number of the two-level systems is restricted to one. 

\begin{figure}
        \centering
        \includegraphics[width=0.48\textwidth]{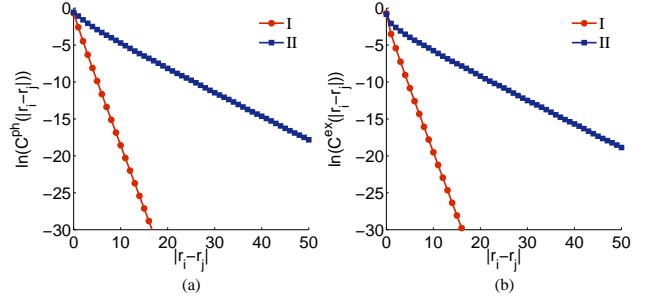}
        \caption{(Color online) Correlation function \fc{a} for photons and \fc{b} for two-level excitations in the first Mott lobe. The Roman numerals in the legend refer to the parameters marked in \figc{fig:pd}{b}. }
        \label{fig:cLobe1}
\end{figure}

In order to display the slowing down of correlations upon
  approaching the boundary of the Mott phase, we evaluate the
spatial correlation function $C^x(|\ve{r}_i-\ve{r}_j|)$ in the first Mott lobe (\fig{fig:cLobe1}).
%fig:cLobe1
The spatial correlation function can be obtained from the Fourier transform of the momentum
distribution. 
For small distances  $|\ve{r}_i-\ve{r}_j|$ between sites $i$ and $j$ the correlation function is a superposition of multiple exponential functions with distinct strengths of decay. For large distances, however, the exponential function with the smallest decay dominates and thus the correlation function is of the form
\begin{equation}
 C^x(|\ve{r}_i-\ve{r}_j|) \propto e^{-\alpha^x \, |\ve{r}_i-\ve{r}_j|} \;\mbox{,}
\end{equation}
as expected in the insulating phase.
Using VCA we are able to extract the correlation
length $\xi^x = 1/\alpha^x$, as data are available for large distances
between two sites $i$ and $j$. From a linear fit for sufficiently
large distances we obtain $\alpha_{\mbox{\footnotesize{I}}}^{ph} =
\alpha_{\mbox{\footnotesize{I}}}^{ex} = 1.711 \pm 0.001$ for the
parameters I, see marks in \figc{fig:pd}{b}, and
$\alpha_{\mbox{\footnotesize{II}}}^{ph} =
\alpha_{\mbox{\footnotesize{II}}}^{ex} = 0.317 \pm 0.001$ for the
parameters II. 
Therefore,
 the slope of the correlation function is 
the same for the two
particle
 species, which is due to the coupling between the photons
 and the two-level excitations. 
As in the BH model \cite{teichmann_process-chain_2009} the absolute
slope $\alpha^x$ of the correlation function shrinks with increasing
hopping strength, which is a precursor of the superfluid phase, where
the correlation between sites persists up to long distances.

\subsection{Nonzero detuning}
\label{nonz}
\begin{figure}
        \centering
        \includegraphics[width=0.48\textwidth]{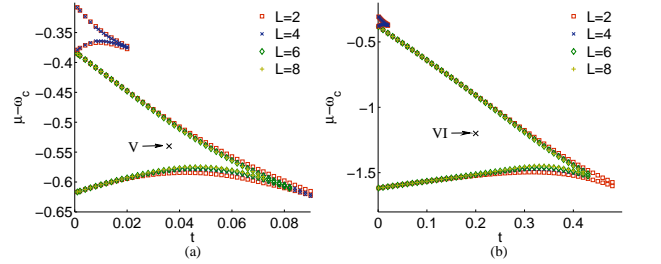}
        \caption{(Color online) Phase boundaries of the 1D JCL model for the detuning \fc{a} $\Delta=-1$ and \fc{b} $\Delta=1$. The marks refer to the parameters where spectral functions are evaluated. }
        \label{fig:pdDet}
\end{figure}

\begin{figure}
        \centering
        \includegraphics[width=0.48\textwidth]{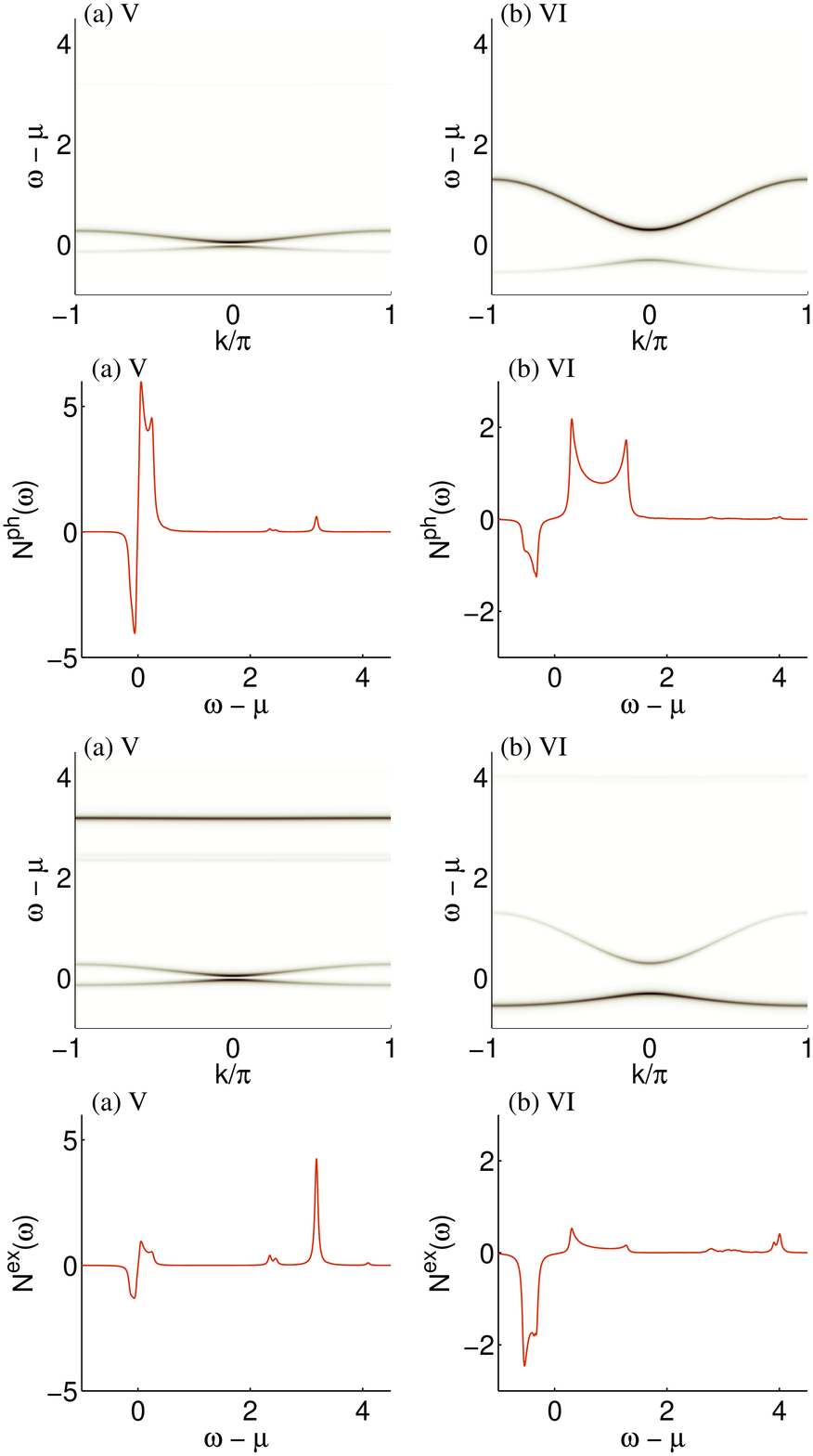}
        \caption{(Color online) Photon spectral function $A^{ph}(\ve{k},\,\omega)$, first row, and density of states $N^{ph}(\omega)$, second row. Two-level excitation spectral function $A^{ex}(\ve{k},\,\omega)$, third row, and density of states $N^{ex}(\omega)$, fourth row. The spectral functions are evaluated for the parameters \fc{a} $t=0.036$, $\mu-\omega_c=-0.54$, $\Delta=-1$ and \fc {b} $t=0.2$, $\mu-\omega_c=-1.2$, $\Delta=1$. The Roman numerals in the captions of the subfigures refer to the marks in \fig{fig:pdDet}. }
        \label{fig:sfDet}
\end{figure}

\begin{figure}
        \centering
        \includegraphics[width=0.48\textwidth]{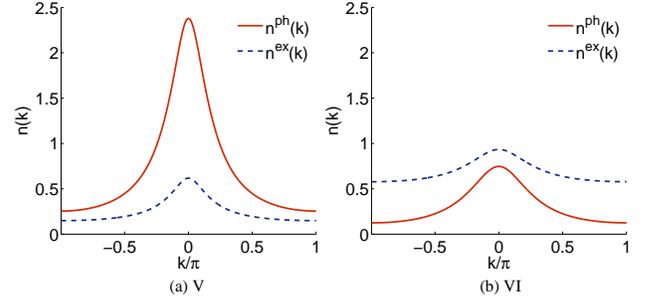}
        \caption{(Color online) Momentum distribution evaluated for the parameters marked in \fig{fig:pdDet}, where \fc{a} corresponds to the parameters V, \ie, negative detuning $\Delta = -1$ and \fc{b} to the parameters VI, \ie, positive detuning $\Delta=1$.}
        \label{fig:nDet}
\end{figure}

The detuning $\Delta$, which is the difference between the resonance frequency $\omega_c$ of the cavities and the energy spacing $\epsilon$ of the two-level systems, is a very important parameter of the JCL model. By varying the detuning it is possible to change the width of the Mott lobes. Phase boundaries obtained by means of VCA with the set of variational parameters $\mat{x} = \lbrace \omega_c,\,\epsilon \rbrace$ for $\Delta=-1$ and $\Delta=1$ are shown in \fig{fig:pdDet}.
%fig:pdDet
For the parameters marked with $\mat{x}$ we evaluate the spectral function of photons $A^{ph}(\ve{k},\,\omega)$ and two-level excitations $A^{ex}(\ve{k},\,\omega)$, see \fig{fig:sfDet}.
%fig:sfDet
An interesting effect can be observed in the spectral functions $A^{ex}(\ve{k},\,\omega)$. Namely, the intensity of the upper band $\omega_p^+$ depends significantly on the detuning $\Delta$. For negative detuning $\Delta=-1$, the upper mode in $A^{ex}(\ve{k},\,\omega)$ is very intense, see \figc{fig:sfDet}{a}, whereas it is almost not visible for positive detuning $\Delta=1$. This behavior remains valid when the spectral functions for positive and negative detuning are evaluated for identical hopping strength. The zero-hopping result for the energy of the upper mode is $\omega_{p,1}^+ \approx 3.15$ for the spectral function shown in \figc{fig:sfDet}{a} and $\omega_{p,1}^+ \approx 3.82$ for the spectral function shown in \figc{fig:sfDet}{b}. The momentum distributions of photons $n^{ph}(\ve{k})$ and two-level excitations $n^{ex}(\ve{k})$ for the parameters marked in \fig{fig:pdDet} are shown in \fig{fig:nDet}.
%fig:nDet
For negative detuning it is energetically more expensive to excite the two-level system than to add a photon to the cavity. Thus, the momentum distribution of photons $n^{ph}(\ve{k})$ dominates over the momentum distribution of two-level excitations $n^{ex}(\ve{k})$. For positive detuning the situation is reversed and  $n^{ex}(\ve{k})$ is larger than  $n^{ph}(\ve{k})$ for all values of the momentum.

\subsection{Polariton quasiparticles}
\label{pola}
Up to now we investigated the photon properties and the two-level excitation properties of the JCL model separately, by extracting the Green's function of photons $G^{ph}(\ve{k},\,\omega) = G_{a_\ve{k}a_\ve{k}^\dagger}(\omega)$ and the Green's function of two-level excitations $G^{ex}(\ve{k},\,\omega) = G_{\sigma^-_\ve{k}\sigma_\ve{k}^+}(\omega)$ from the compound Green's function $\mat{G}(\ve{k},\,\omega)$, which is a $2\times2$ matrix of the form
\begin{equation}
 \mat{G}(\ve{k},\,\omega) = \left(
\begin{array}{cc}
 G_{a_\ve{k}a_\ve{k}^\dagger}(\omega) & G_{a_\ve{k}\sigma^+_\ve{k}}(\omega) \\
 G_{\sigma^-_\ve{k} a_\ve{k}^\dagger}(\omega) & G_{\sigma^-_\ve{k}\sigma_\ve{k}^+}(\omega) \\
\end{array}
 \right) \; \mbox{.}
 \label{eq:gf2by2}
\end{equation}

\begin{figure}
        \centering
        \includegraphics[width=0.48\textwidth]{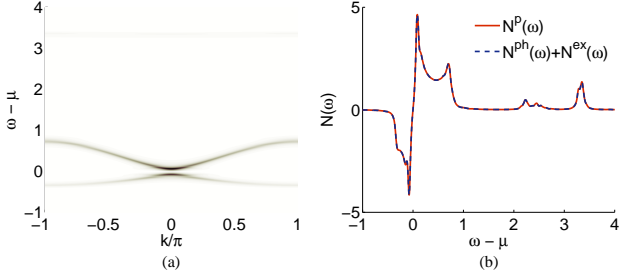}
        \caption{(Color online) Polariton spectral function \fc{a} and density of states \fc{b} evaluated for the parameters II corresponding to the first Mott lobe, \ie, $t=0.12$, $\mu-\omega_c=-0.84$, and $\Delta=0 $. In \fc{b} the polariton density of states $N^p(\omega)$ is compared with the sum of the photon density of states $N^{ph}(\ve{k},\,\omega)$ and the two-level excitation density of states $N^{ex}(\ve{k},\, \omega)$, which coincide by definition. }
        \label{fig:sfDosPolLobe1}
\end{figure}

\begin{figure}
        \centering
        \includegraphics[width=0.48\textwidth]{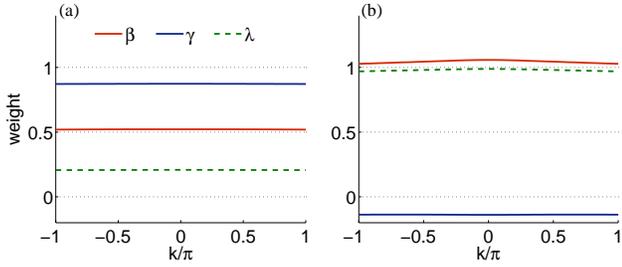}
        \caption{(Color online) Photon contribution $\beta$ and two-level excitation contribution $\gamma$ to the polariton quasiparticles. \fc{a} shows the results for $p_{+,\ve{k}}^\dagger$ corresponding to the upper particle band $\omega_p^+$ and \fc{b} for $p_{-, \ve{k}}^\dagger$ corresponding the lower particle band $\omega_p^-$.
Additionally to the weights $\beta$ and $\gamma$ the overlap $\lambda$ is shown. }
        \label{fig:qwWeightssLobe1}
\end{figure}

Next we will discuss the polaritonic properties of the JCL model.
We start out with the first Mott lobe for zero detuning and focus again on the
parameter set marked as II in \fig{fig:pd}, \ie, $t=0.12$, $\mu-\omega_c=-0.84$ and $\Delta=0$.
The polaritonic spectral function
$A^p(\ve{k},\,\omega)$ and the corresponding density of states $N^p(\omega)$, which is by construction identical to the total density of states of photons plus two-level excitations, is shown in \fig{fig:sfDosPolLobe1}.
%fig:sfDosPolLobe1
For the first Mott lobe the hole case is special since both, $\sigma^- \ket{n,-} \propto \ket{0,-}$
and $a \ket{n,-} \propto \ket{0,-}$ yield  the exact zero-particle state. Consequently, the polariton can be chosen ad libitum, it will always be exact. Therefore
in \fig{fig:qwWeightssLobe1} only the particle part of the polaritonic weights is depicted.
%fig:qwWeightssLobe1
The right panel represents the result for the lower particle excitation. The polariton has very pronounced photonic character and the weights of photons and two-level system have opposite sign.
Interestingly, the lower particle excitation can very well be mimicked by a single polariton on top of the $N_p$-particle ground state, as can be inferred from the fact that $\lambda\approx 1$.
Moreover, a slight $\ve{k}$-dependence of the weights is observed.
Contrarily in the upper particle band, the polariton has pronounced two-level-system character, the weights have the same sign, there is almost no $\ve k$-dependence, and the polariton description is poor ($\lambda\approx 0.2$).

\begin{figure}
        \centering
        \includegraphics[width=0.48\textwidth]{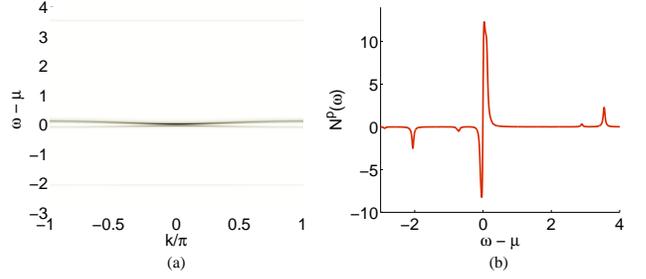}
        \caption{(Color online) Polariton spectral function \fc{a} and density of states \fc{b} evaluated for the parameters IV corresponding to the second Mott lobe, \ie, $t=0.012$, $\mu-\omega_c=-0.38$, and $\Delta=0$. }
        \label{fig:sfPol}
\end{figure}

\begin{figure}
        \centering
        \includegraphics[width=0.48\textwidth]{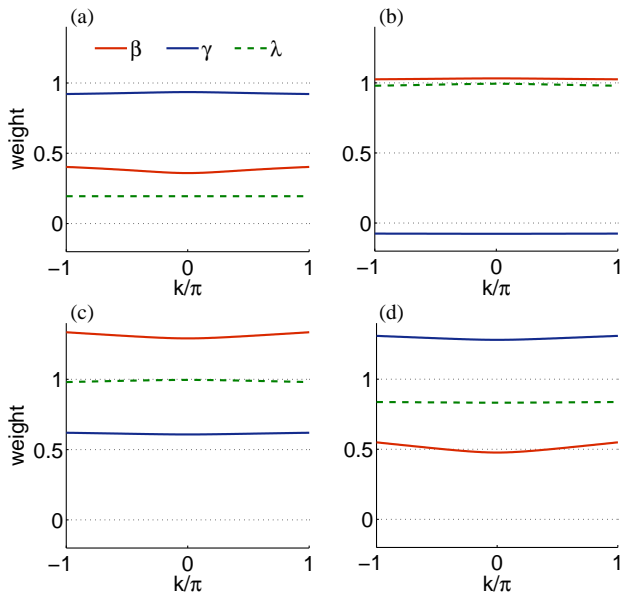}
        \caption{(Color online) Photon contribution $\beta$ and two-level excitation contribution $\gamma$ to the polariton quasiparticles. \fc{a} shows the results for $p_{+,\ve{k}}^\dagger$ corresponding to the upper particle band $\omega_p^+$, \fc{b} for $p_{-,\ve{k}}^\dagger$ corresponding the lower particle band $\omega_p^-$, \fc{c} for $h_{-,\ve{k}}^\dagger$ corresponding to the lower hole band $\omega_h^-$ and \fc{d} for $h_{+,\ve{k}}^\dagger$ corresponding to the upper hole band $\omega_h^+$. Additionally to the weights $\beta$ and $\gamma$ the overlap $\lambda$ is shown. }
        \label{fig:qwBands}
\end{figure}

Now we turn to the second Mott lobe, which allows us to study the hole polariton as well.
The polariton spectral function and the corresponding density of states  evaluated for the parameters IV, \ie, $t=0.012$, $\mu-\omega_c=-0.38$, and $\Delta=0$, are shown in \fig{fig:sfPol}.
%fig:sfPol
The weights are shown along with the overlap $\lambda$ in \fig{fig:qwBands}.
%fig:qwBands
The lower bands $\omega_{p/h}^-$ are well described by the quasiparticles as the overlap $\lambda$ is almost one for both bands. The upper bands $\omega_{p/h}^+$, however, are not described that well. In particular $\lambda \approx 0.2$ for the upper particle band and $\lambda \approx 0.85$ for the upper hole band. The weights $\beta$ and $\gamma$ are significantly more wave vector dependent, especially for the  upper bands $\omega_{p/h}^+$, \ie, $\alpha=+$.
Apart from the more pronounced $\ve{k}$-dependence, the weights for the particle case are rather similar to those of  the first Mott lobe.
However, there are striking differences in the weights for the particle and hole part within the second Mott lobe.
First, the $\ve{k}$ dependence is  more pronounced. Second, the sign of the relative weights is positive for both bands $\alpha=\pm$, and finally,
the composition of the polariton in the two bands is reverse. The lower band has predominantly photonic character, while opposite holds for the upper band.

Eventually, we want to compare the VCA results with those of the single-site problem, which are derived in \app~\ref{app:d}. In the single-site problem the sign of the relative polaritonic weights is the same as that observed in the lattice. In the first Mott lobe  the relative weights for the particle case are for the upper band
$q_+\equiv {\gamma^+_{p,n=1}}/{\beta^+_{p,n=1}}=\sqrt{2}+1$ and for the lower band the reciprocal relation holds $q_-\equiv{\beta^-_{p,n=1}}/{\gamma^-_{p,n=1}}=-q_+$.
There is agreement in the relative signs and the composition of the polariton between the single-site limit and the lattice system, but the reciprocal property is strongly violated
in the lattice case. This might be understood as follows. The
itinerant particles are the photons. In order to gain kinetic energy
it is 
convenient
for the system to
increase the photonic character in the dispersive lower band, depicted in \figc{fig:sfPol}{a}. The upper band, on the other hand, has little dispersion and behaves more like the single-site limit.

In the second Mott lobe, the relative weights obtained in the single-site limit for particle excitations are
$q_+ =\sqrt{3}+\sqrt{2}$ and $q_- = -q_+$. Like in the lattice case, the weights of the particle part are comparable in the first and second Mott lobe. Quantitatively, the relative weight $|q_\pm|$ is roughly $30\%$ larger in the second Mott lobe, which is also the case in the lattice system. As far as the hole part is concerned, the single-site limit nicely corroborates all observations of the lattice model.

In the single-site problem, the exact many-body eigenstates $\ket{n\pm 1,\alpha}$ can be generated exactly by suitable polariton operators
acting on the state $\ket{n,-}$. This is no longer the case in the lattice due to local particle number fluctuations induced by particle motion.
Already in the single-site limit, the polariton operators are, however, not universal, they depend on the filling $n$ and in the lattice case even on the wave vector $\ve k$. On top of that, the polariton operator for holes is not the adjoint of the corresponding polariton  creation operator of the particle type, or in other word its annihilation operator.

\section{\label{sec:conclusion}Conclusions}
In this paper we presented and discussed the spectral properties of the Jaynes-Cummings lattice model in one dimension obtained within the variational cluster approach. Using the resonance frequency $\omega_c$ of the cavities and the energy spacing $\epsilon$ of the two-level systems
as variational parameters
in the variational cluster approach procedure
provides a significant improvement with respect to 
the case of a single variational parameter.
On the one hand, varying both $\omega_c$ as well as $\epsilon$ 
(or, at least $\mu$)
is necessary to achieve a correct particle density in the original system and on the other hand improved results for the phase boundaries, and thus, for the spectral functions as well, are obtained due to the augmented set of variational parameters. In order to apply the variational cluster approach and include $\epsilon$ as variational parameter the two-level systems have been mapped onto hard-core bosons, which yields correct poles of the Green's function in the relevant energy range.
We evaluated and discussed spectral functions for photons and two-level excitations. The spectral functions generally consist of four bands, cosinelike shaped lower particle/hole bands, which are centered around zero energy, and essentially flat upper particle/hole bands. An exception are the spectral functions in the first Mott lobe, which contain the two lower bands but
from the upper bands only the particle part.
Using first-order degenerate perturbation theory, we evaluated
analytical expressions for the bands, which allowed us to explain why
the upper modes are essentially flat whereas the lower modes exhibit a
pronounced cosinelike shape. Additionally, we compared the analytical
solution for the bands with the variational cluster approach results. For small
  hopping strength $t$ we observe, as expected, good agreement between
  the two approaches. However for parameters located close to the tip
  of 
the Mott lobe,
  first-order degenerate perturbation theory yields results that differ from the exact ones in  both,  shape and  width of the bands.
Furthermore, 
we evaluated densities of states, momentum distributions and spatial correlation functions for photons and two-level excitations. We also investigated detuning effects on the spectral properties and found indications that the intensity of the upper particle band of the two-level excitation spectral function depends strongly on the detuning. Based on the information obtained from the photons and two-level excitations we investigated the polaritonic properties of the Jaynes-Cummings lattice model. Therefore we introduced wave vector and filling dependent polariton particle creation and hole creation operators, which are linear combinations of photon and two-level excitation creation operators. We evaluated spectral functions and densities of states based on the polariton quasiparticles and analyzed the weights of their constituents. We have seen that the polariton operators are nontrivial combinations of photon and two-level system operators, which depend on the wave vector, the quasiparticle band, and the filling, or rather the Mott lobe. On top of that, the  polariton operators of particle and hole type are not adjoint operators. It is therefore not possible to describe the JCL model by a simple single-band polariton model.

\appendix

\section{\label{app:c} Properties of the VCA Green's function}

Here, we will derive some properties of the VCA 
Green's function 
 for bosons. To begin with, we define
new operators $d^\dagger_{J,r,\tilde{\ve k}}$ with $d^\dagger_{1,r,\tilde{\ve k}} \equiv a^\dagger_{r,\tilde{\ve k}}$ and $d^\dagger_{2,r,\tilde{\ve k}} \equiv \sigma^+_{r,\tilde{\ve k}}$, where $r$ stands for the site number within the clusters and $\tilde{\ve k}$ is the wave vector of the first Brillouin zone of the superlattice.
The VCA Green's function will still be diagonal in the latter index due to the periodicity of the clusters.
The spectral representation of the cluster Green's function
\[\mat{G}^\prime_{IJ}(\tilde{\ve k},\,\omega)\equiv
\ll d_{I,r,\tilde{\ve k}};d_{J,s,\tilde{\ve k}}^\dagger \gg_\omega\]
can  be written in the compact form using the so-called $\mat{Q}$-matrices \cite{aichhorn_variational_2006,knap_spectral_2010}
\begin{align*}
    \mat{G}^\prime &= \mat{Q} \, \mat{D}^\prime_\omega \mat{S} \, \mat{Q}^\dagger\;.
\end{align*}
Here, $\mat{Q}$ is a $M \times K$ matrix, where $M$ is twice the number of cluster sites (the factor 2 stems from the two species of operators) and $K=K_p+K_h$, where $K_p$ and $K_h$ is the dimension of the Hilbert space for $N_p + 1$ and $N_p - 1$ particles, respectively. The $\mat{Q}$-matrix is defined as follows
\begin{align*}
        Q_{I,r;\nu} &=
    \begin{cases}
    \bra{\psi_0} d_{I,r,\ve k} \ket{\psi^{N_p+1}_\nu}&\text{for } \nu \le K_p\\
    \bra{\psi_0} d_{I,r,\ve k}^\dagger \ket{\psi^{N_p-1}_{\nu-K_p}}&\text{for } \nu > K_p\;
    \end{cases} \; \mbox{.}
\end{align*}
The diagonal matrix $\mat{D}^\prime_\omega=\diag(\omega-\omega^\prime_\nu)^{-1}$  contains the individual poles $\omega^\prime_\nu$ of the cluster and $\mat{S}=\diag(s_\nu)$ is a diagonal sign matrix with  $s_\nu =+1$ for particle
excitations ($\omega_\nu^\prime>0$) and $s_\nu=-1$ for hole excitations
($\omega_\nu^\prime<0$).
The VCA Green's function in $\mat{Q}$-matrix representation for bosons \cite{knap_spectral_2010} reads
\begin{align}
    \mat{G}(\tilde{\ve k},\omega) &=
     \mat{Q} \, \mat{X} \, \mat{D}_\omega \,
    \mat {X}^{-1} \, \mat{S} \, \mat{Q}^\dagger  \;,
\end{align}
where $\mat{D}_\omega=\diag ({\omega-\omega_\nu})^{-1}$ is the diagonal matrix of the individual poles at the VCA energies. These energies
and the corresponding eigenvector matrix $\mat{X}$ are determined via the generalized eigenvalue problem
\begin{align*}
    \underbrace{(
    \diag(s_\nu \, \omega^\prime_\nu)  + \mat{Q}^\dagger \, \mat{V} \, \mat{Q} )}_{\equiv \mat{M}} \mat{X}
     = \mat{S}\, \mat{X} \mat{\Delta}\;\mbox{,}
\end{align*}
where $\mat{V}=\mat{H}_0-\mat{H}_0^\prime$ is the difference of the matrices of the single-particle part of the Hamiltonian for the original and the reference system (\ie, the cluster).

A general feature of such eigenvalue equations for Hermitian matrices $\mat M$ is that both $ \mat{X}^\dagger \, \mat{M} \, \mat{X}$ and $\mat{X}^\dagger \, \mat{S} \, \mat{X} \equiv \diag(\kappa_\nu^{-1})$ are diagonalized, but $\mat X$ is not unitary.
We can exploit this fact as follows
\begin{align*}
    \mat{G}(\tilde{\ve k},\, \omega) &=
\mat Q \, \mat X \, \mat{D}_\omega \, \underbrace{\mat{X}^{-1} \, \mat{S} \, (\mat{X}^{-1})^\dagger}_{\equiv \mat{D}_\kappa} \mat{X}^\dagger \, \mat{Q}^\dagger    \;\mbox{.}
\end{align*}
The matrix $\mat{D}_\kappa = (\mat{X}^\dagger \, \mat{S} \, \mat{X})^{-1}$ is diagonal as we just discussed and can be combined with  $\mat{D}_\omega$ resulting in
\begin{align*}
    \mat{G}(\tilde{\ve k},\,\omega) &=
\mat{Q} \, \mat{X} \, \tilde{\mat{D}}_\omega \, (\mat{Q}\, \mat{X})^\dagger  \\
( \tilde{\mat{D}}_\omega )_{\nu\nu'} &= \delta_{\nu,\nu'}\;\frac{\kappa_\nu}{\omega-\omega_\nu}\;.
\end{align*}
Moreover, the pole strengths $\kappa_\nu$ are real since 
\begin{align*}
    \kappa_\nu^{-1} &= (\mat X^\dagger\, \mat S \, \mat{X})_{\nu\nu}\\
    &=\sum_\mu s_\mu \, |X_{\nu\mu}|^2\;\mbox{.} 
\end{align*}
When the VCA parameters are determined consistently, the stability of the $N_p$-particle system requires that the sign of $\kappa_\nu$ coincides with the sign of the excitation energies $\omega_\nu$, like in the exact spectral representation.

So far the Green's function still depends on the intra cluster indices $r,\,s$. The purely $\ve k$-dependent Green's function is commonly obtained by Green's function-periodization.\cite{snchal_spectral_2000, snchal_introduction_2008}
Invoking the periodization prescription yields the Green's function matrix merely in the indices $I,\,J$ for the two particle species
\begin{align}
    \mat{G}(\ve k,\, \omega) &=
   \tilde{\mat{Q}} \, \mat{X} \, \tilde{\mat{D}}_\omega \, \mat{X}^\dagger \, \tilde{\mat{Q}}^\dagger\label{eq:VCAxxx}\\
    \text{with}\quad
   \tilde{Q}_{I,\nu} &=
\frac{1}{N}\sum_r e^{-i \, \ve k \, \ve x_r} Q_{I,r; \, \nu}\;\mbox{,}
\end{align}
see also \eq{eq:gf2by2}. \Eq{eq:VCAxxx} corresponds to the spectral representation of the exact Green's function and it allows to extract the VCA approximation of the many-body eigenstates of the infinite system, which are obviously a linear combination of the cluster eigenstates for both, particle and hole excitations.

As described in the text we need the integrated spectral density, \ie,
\begin{align*}
    A_{IJ}(\ve k, \, \Omega_\alpha) &= \int_{\Omega_\alpha} \; (
    -\frac{1}{\pi} \text{Im} \, \;G_{IJ}(\ve k,\,\omega + i \eta)
    )\;d\omega\\
    &= \sum_{\nu, \, \omega_\nu(\ve k)\in \Omega_\alpha}
    (\tilde{\mat Q}\, \mat{X} )_{I,\nu} \, \kappa_\nu \,
        ( \tilde{\mat Q} \, \mat{X})_{J,\nu}\;.
\end{align*}
We readily recognize, that the integrated spectral density is either positive or negative definite, depending on whether the quasiparticle under consideration is of particle or hole type.
Equivalently, in the original representation
\begin{align*}
    A_{IJ}(\ve k,\,\Omega_\alpha)
    &= \sum_{\nu, \, \omega_\nu(\ve k)\in \Omega_\alpha}
    (\tilde{\mat{Q}} \, \mat{X})_{I,\nu}
        (\mat{X}^{-1} \mat{S} \, \tilde{\mat{Q}}^\dagger)_{\nu,J}\;\mbox{.}
\end{align*}
For the polariton discussion it is 
convenient
to suppress the minus sign arising in the hole case and we define the strictly positive integrated spectral densities as
\begin{align}
    \tilde{ A}_{IJ}(\ve k,\, \Omega_\alpha) \equiv | A_{IJ}(\ve k,\, \Omega_\alpha) |\;\mbox{.}
\end{align}

\section{\label{app:a}Solution of the single-site problem}
For zero-hopping strength $t=0$ the JCL model can be solved exactly, as it reduces to a single-site problem, \ie, to the JC model. Including the chemical potential yields the single-site Hamiltonian
\begin{equation}
  \hat{H}^{JCL}_S = \hat{H}^{JC} - \mu\,(a^\dagger\,a+\sigma^+\sigma^-) \; \mbox{,}
 \label{eq:jcsinglesite}
\end{equation}
where we dropped the site index $i$. It can be evaluated with respect to the bare states $\ket{n_p,\,s}$, where $n_p$ is the number of photons and $s \in \lbrace \downarrow,\,\uparrow \rbrace$.
Next, we sketch the most important steps for solving the single-site JCL model. A detailed discussion can be found for example in Refs.~\onlinecite{haroche_exploringquantum:_2006} or \onlinecite{hussin_ladder_2005}.
As the JC Hamiltonian conserves the particle number the Hamiltonian $\hat{H}^{JCL}_S$ is block diagonal. Each block corresponds to a certain particle number $n$ and thus we use the bare states $\ket{n-1,\,\uparrow}$ and $\ket{n,\,\downarrow}$ to evaluate the block, which yields
\begin{equation}
 B_n=\left( \begin{array}{cc}
         (n-1)\omega_c + \epsilon -\mu\,n & \sqrt{n} \\
         \sqrt{n} & n\,\omega_c -\mu\,n
        \end{array}
 \right) \;\mbox{,}
 \label{eq:jcsinglesiteblock}
\end{equation}
when using as denoted in \se~\ref{sec:model} the coupling $g$ as unit of energy. The eigenvalues of the block $B_n$ are
\begin{equation}
 E_{\ket{n,\,\alpha}} = n\,\omega_c - \frac{\Delta}{2} + \alpha \, q(n) - \mu\,n \;\mbox{,}
 \label{eq:jcsinglesiteblockeval}
\end{equation}
where $\alpha \in \lbrace -,\,+ \rbrace$ and $q(n)=\sqrt{\left({\Delta}/{2}\right)^2 + n}$. For a certain particle number $n$ the energy $E_{\ket{n,\,-}}$ is always smaller than $E_{\ket{n,\,+}}$ and thus $E_{\ket{n,\,-}}$ is the ground state energy of the sector with $n$ particles, \ie, of the block $B_n$. The eigenvectors $\ket{n,\,\alpha}$ of the matrix $B_n$ are termed dressed states and are given by
\begin{equation}
 \ket{n,\,\alpha} = u_{n\alpha} \ket{n-1, \,\uparrow} + v_{n\alpha} \ket{n, \,\downarrow}\;\mbox{,}
 \label{eq:dressedstates}
\end{equation}
where $n>0$, $(u_{n+},\,v_{n+}) \equiv (\sin \theta(n),\, \cos
\theta(n))$ and $(u_{n-},\,v_{n-}) \equiv (\cos \theta(n),\, -\sin
\theta(n))$ with the following relations $\sin
\theta(n)=\sqrt{({q(n)-\Delta/2})/{2q(n)}}$ and $\cos
\theta(n)=\sqrt{({q(n)+\Delta/2})/{2q(n)}}$. An exception is the bare
state $\ket{0,\,\downarrow}$, which forms a $1\times1$ block of zero
particles and has the eigenvalue $E_{\ket{0,\,\downarrow}}=0$,
independently of the detuning $\Delta$. According to the notation used
in 
\eq{eq:dressedstates}, we denote this state as $\ket{0,\,-}$. In order
to obtain the phase boundary for zero hopping between two adjacent
Mott 
lobes, the energies $E_{\ket{n,\,-}}$ and $E_{\ket{n+1,\,-}}$ have to be set equal.
The energies of the states $\ket{m,\,-}$ are used, as the phase diagram is evaluated for the ground state.
The comparison of the energies yields $(\mu - \omega_c) = q(n) - q(n+1)$ for the location of the phase boundary at zero hopping.

\section{\label{app:b}First-order degenerate perturbation theory}
In this appendix we evaluate the results of first-order degenerate perturbation theory for the single-particle and single-hole excitation bands of the JCL model.
To apply first-order degenerate perturbation theory the matrix elements of the perturbation $\hat{H}_1 = \sum_{ij} t_{ij} \, a_i^\dagger \, a_j$, where $t_{ij}$ is the hopping matrix, have to be evaluated with respect to the degenerate states $\ket{\Psi_p^{\alpha,\,l}}$ and $\ket{\Psi_h^{\alpha,\,l}}$, see \eq{eq:psiDegenerate}. As the hopping term $\hat{H}_1$ does not change the total particle number and does not effect the excitation $\alpha$, the following two matrices have to be evaluated; one for single-particle excitations
\begin{equation}
 (\mat{M}_p^\alpha)_{ll^\prime} \equiv \langle \Psi_p^{\alpha,\,l} | \hat{H}_1 | \Psi_p^{\alpha,\,l^\prime} \rangle
 \label{eq:mp}
\end{equation}
and one for single-hole excitations
\begin{equation}
 (\mat{M}_h^\alpha)_{ll^\prime} \equiv \langle \Psi_h^{\alpha,\,l} | \hat{H}_1 | \Psi_h^{\alpha,\,l^\prime} \rangle \;\mbox{.}
 \label{eq:mh}
\end{equation}
Plugging \eq{eq:psi1} in \eq{eq:mp} yields
\begin{align}
 (\mat{M}_p^\alpha)_{ll^\prime} = \bigotimes_{ \genfrac{}{}{0pt}{}{\nu = 1}{\nu \neq l} }^N \bra{n,\,-}_\nu \bra{n+1,\,\alpha}_l \sum_{i,\,j} t_{ij} \, a_i^\dagger \, a_j& \nonumber \\
 \ket{n+1,\,\alpha}_{l^\prime} \bigotimes_{  \genfrac{}{}{0pt}{}{\nu^\prime = 1}{\nu \neq l^\prime} }^N \ket{n,\,-}_{\nu^\prime}&  \;  \mbox{.}
 \label{eq:mp2}
\end{align}
Due to the orthogonality of the eigenvectors of sectors with different particle number, the conditions $i=l$ and $j=l^\prime$ hold, which reduce the matrix elements to
\begin{align}
 (\mat{M}_p^\alpha)_{ll^\prime} &= t_{ll^\prime} \bra{n,\,-}_{l^\prime} \bra{n+1,\,\alpha}_l a_l^\dagger \, a_{l^\prime}  \ket{n+1,\,\alpha}_{l^\prime}\ket{n,\,-}_l \nonumber \\
 &= t_{ll^\prime} | \bra{n+1,\,\alpha} a^\dagger \ket{n,\,-} |^2 \;  \mbox{.}
 \label{eq:mp3}
\end{align}
In the second step, we dropped the site index as the expectation value does not depend on the specific lattice site. The corrected matrix elements are thus the old ones with renormalized hopping strength
\begin{align}
 -\tilde{t}_p^\alpha &\equiv -t | \bra{n+1,\,\alpha} a^\dagger \ket{n,\,-} |^2  \nonumber \\
 &= -t |\sqrt{n} \, u_{n+1\alpha} \, u_{n-} + \sqrt{n+1} \, v_{n+1\alpha} \, v_{n-} |^2 \;\mbox {.}
 \label{eq:teffp}
\end{align}
Analogously, one obtains
\begin{equation}
 (\mat{M}_h^\alpha)_{ll^\prime} = t_{l^\prime l} | \bra{n-1,\,\alpha} a \ket{n,\,-} |^2
\end{equation}
for the matrix elements defined in \eq{eq:mh}. From that the renormalized hopping strength for single-hole excitations is evaluated as
\begin{equation}
 -\tilde{t}_h^\alpha = -t |\sqrt{n-1} \, u_{n-1\alpha} \, u_{n-} + \sqrt{n} \, v_{n-1\alpha} \, v_{n-} |^2 \;\mbox {.}
 \label{eq:teffh}
\end{equation}
The eigenvalues of the matrices $\mat{M}_{p/h}^\alpha$ are the first-order corrections and thus the corrected energies $\mathcal{E}_{\ket{n\pm1,\,\alpha}}(k)$ of the one-dimensional JCL model are given by
\begin{subequations}
\begin{align}
 \mathcal{E}_{\ket{n+1,\,\alpha}}(k) &= E_{\ket{n+1,\,\alpha}} - 2 \, \tilde{t}_p^\alpha \cos k \;\mbox{and} \\
 \mathcal{E}_{\ket{n-1,\,\alpha}}(k) &= E_{\ket{n-1,\,\alpha}} - 2 \, \tilde{t}_h^\alpha \cos k \;\mbox{,}
\end{align}
\end{subequations}
respectively, where $k$ is a wave vector of the first Brillouin zone. Within first-order degenerate perturbation theory we obtain
\begin{align}
 \omega_{p,1}^\alpha &= \mathcal{E}_{\ket{n+1,\,\alpha}}(k) - E_{\ket{n,\,-}} \nonumber \\
  &= (\omega_c - \mu)+\alpha\,q(n+1)+q(n) - 2 \, \tilde{t}_p^\alpha \cos k
  \label{eq:omParticleApp}
\end{align}
for the single-particle excitation band and
\begin{align}
 \omega_{h,1}^\alpha &= E_{\ket{n,\,-}} - \mathcal{E}_{\ket{n-1,\,\alpha}}(k) \nonumber \\
  &= (\omega_c - \mu)-\alpha\,q(n-1)-q(n) + 2 \, \tilde{t}_h^\alpha \cos k \;\mbox{.}
  \label{eq:omHoleApp}
\end{align}
for the singe-hole excitation band.

\section{Polariton operators in the single-site limit\label{app:d}}
In this appendix, we want to analyze the polaritonic feature in the single-site limit for zero detuning.

In the single-site limit it is exactly possible to construct a polariton operator which, applied to the many-body eigenstate $\ket{n,-}$, generates the eigenstates $\ket{n\pm 1,\alpha}$.
The polaritonic weights follow from
\begin{align}
(   \beta a^\dagger &+ \gamma \sigma^+ )\ket{n,-} = \nonumber\\
&=\frac{\beta \sqrt{n} - \gamma}{\sqrt{2}}\ket{n,\uparrow} - \beta\sqrt{\frac{n+1}{2}}\ket{n+1,\downarrow}\nonumber\\
&\overset{!}{=} \ket{n+1,\alpha}\nonumber\\
\frac{\beta^\alpha_{p,n}}{\gamma^\alpha_{p,n}}
&=\frac{1}{\alpha \sqrt{n+1}+  \sqrt{n}}\label{eq:appyy}
\end{align}
Here, we explicitly include the filling $n$ as index. So the relative weights are $(\sqrt{n+1}+\sqrt{n})^{-1}$ for the upper band ($\alpha=+$) and $ - (\sqrt{n+1}+\sqrt{n})$ for the lower band ($\alpha=-$). This means that in the lower band the weights have opposite sign and the polaritons are of predominant photonic character, while the opposite applies to the upper band. The modulus of relative weight is just the inverse, \ie, $|{\beta^+}/{\gamma^+}| = |{\gamma^-}/{\beta^-}|$.

Next, we study the hole case for $n>1$ 
\begin{align}
(   \beta a &+ \gamma \sigma^- )\ket{n,-} = \nonumber\\
&=\beta \sqrt{\frac{n-1}{2}}\ket{n-2,\uparrow} + \frac{ \gamma- \beta\sqrt{n}}{\sqrt{2}}\ket{n-1,\downarrow} \nonumber\\
&\overset{!}{=} \ket{n-1,\alpha}\nonumber\\
\frac{\beta^\alpha_{h,n}}{\gamma^\alpha_{h,n}} &=
\frac{1}{\sqrt{n} + \alpha \sqrt{n-1}}\;,\label{eq:appyy2}
\end{align}
which is positive for both bands $\alpha=\pm$. Again we have the reciprocal property
${\beta^+}/{\gamma^+} = {\gamma^-}/{\beta^-}$ and the lower band has predominantly photonic character, while the opposite is the case in the upper band.

Now we want to scrutinize the generalized eigenvalue problem of the
Green's function. The single-site Green's function reads
\begin{align*}
    G^S_{IJ}(\omega) &= \sum_{\alpha=\pm} \frac{\Qp{I,\alpha} \Qpdag{\alpha,J}}{\omega-\omega_{p,\alpha}} -
    \sum_{\alpha=\pm} \frac{\Qh{I,\alpha} \Qhdag{\alpha,J}}{\omega-\omega_{h,\alpha}} \\
    \Qp{I,\alpha} &= \bra{n+1,\alpha} d_I^\dagger \ket{n,-}^* \\
    \Qh{I,\alpha} &= \bra{n-1,\alpha} d_I \ket{n,-}\;,
\end{align*}
where we introduced the operators $d_1 \equiv a$ and $d_2\equiv \sigma^-$. For the single-particle term we obtain
\begin{align*}
    \Qp{1,\alpha}  &= \bra{n+1,\alpha} a^\dagger \ket{n,-}^*
    = \frac{1}{2}(\sqrt{n} - \alpha \sqrt{n+1})\\
        \Qp{2,\alpha}  &= \bra{n+1,\alpha} \sigma^+ \ket{n,-}^* =
 -\frac12\;.
\end{align*}
With the definition $\ve x_{\alpha} = (\Qp{1,\alpha},\Qp{2,\alpha})^T$ the integrated spectral density  for the particle part can be expressed as
\begin{align*}
    \tilde{\mat A}_\alpha &=  \ve x_{\alpha} \,  {\ve x_\alpha}^T\;.
\end{align*}
The overlap matrix
$\mat{S}_p\equiv\langle d_I d_J^\dagger \rangle$  is readily obtained by the spectral theorem
\begin{align*}
    \mat{S}_p &= \ve x_{+} \,  {\ve x_+}^T  + \ve x_{-} \,  {\ve x_-}^T \;,
\end{align*}
and the generalized eigenvalue problem for the polariton weights according to \eq{eq:gen_ev} reads
\begin{align*}
    (1-\lambda) \, \ve x_{\alpha} \,  {\ve x_\alpha}^T \, \tilde{\ve{z}}_\alpha &= \lambda \, \ve x_{-\alpha} \,  {\ve x_{-\alpha}}^T \, \tilde{\ve{z}}_\alpha\;.
\end{align*}
The eigenvalues are zero and one. For the polariton weights we are interested in the latter. The corresponding eigenvector is simply given by the vector orthogonal to $\ve x_{-\alpha}$ 
\begin{align*}
    \tilde{\ve z}_\alpha &=\frac12
    \begin{pmatrix}
    1\\
    \sqrt{n} + \alpha \sqrt{n+1}
    \end{pmatrix} \;\mbox{.}
\end{align*}
With that one obtains for the ratio of the weights
\begin{align*}
    \frac{\beta^\alpha_{p,n}}{\gamma^\alpha_{p,n}} &= \frac{1}{\sqrt{n} + \alpha \sqrt{n+1}}\;,
\end{align*}
which is in agreement with the exact result in \eq{eq:appyy}.

Now we address the hole case, again for $n>1$,
\begin{align*}
    \Qh{1,\alpha}  &= \bra{n-1,\alpha} a \ket{n,-}
    = \frac{1}{2}(\sqrt{n-1} - \alpha \sqrt{n})\\
        \Qh{2,\alpha}  &= \bra{n+1,\alpha} \sigma^- \ket{n,-} =
 \alpha\frac12\;\mbox{.}
\end{align*}
We proceed as in the particle case with the definition of $\ve x_{\alpha}\,^T = (\Qh{1,\alpha},\Qh{2,\alpha})$. The remaining steps are the same as before and we end up with
\begin{align*}
    \frac{\beta^\alpha_{h,n}}{\gamma^\alpha_{h,n}} &= \frac{\alpha}{\sqrt{n-1} + \alpha \sqrt{n}}
    = \frac{1}{\sqrt{n} + \alpha \sqrt{n-1}}\;\mbox{,}
\end{align*}
which is also in agreement with the exact result, see \eq{eq:appyy2}. So we see that the determination of the polaritonic weight via the generalized eigenvalue problem is reasonable.
In the single-site limit, the exact many-body eigenstates $\ket{n\pm 1,\alpha}$ can be generated correctly  by suitable polariton operators
acting on the state $\ket{n,-}$.
The operators are, however, not universal, they depend on $n$ and in the lattice case even on $\ve k$. On top of that, the polariton creation operator for holes is not the adjoint of the corresponding polariton creation operator of the particle type, or in other words its annihilation operator.

\begin{acknowledgments}
We are grateful to D. Rossini for providing us the DMRG Data of the phase boundaries used in \figg{fig:pd}{fig:comparison}. We made use of parts of the ALPS
library (Ref.~\onlinecite{albuquerque_alps_2007}) for the implementation of lattice geometries and for parameter parsing.
We acknowledge financial support from the Austrian Science Fund (FWF) under the doctoral program ``Numerical Simulations in Technical Sciences'' No. W1208-N18 (M.K.) and under Project No. P18551-N16 (E.A.).
\end{acknowledgments}


\begin{thebibliography}{40}
\bibitem[{har()}]{hardcore}
\bibinfo{note}{This implies, for example, that for some operator products, such
  as $b_i b^{\dagger}_i\not=\sigma_i^- \sigma_i^+$, the mapping is not
  correct.}
\end{thebibliography}
\end{document}